\documentclass[%
 reprint,
 superscriptaddress,
 preprintnumbers,
 amsmath,amssymb,
 aps
]{revtex4-2}

\usepackage[T1]{fontenc} 
\usepackage[utf8]{inputenc}
\usepackage[USenglish]{babel}
\usepackage{multirow}
\usepackage[dvipsnames]{xcolor}
\usepackage{graphicx} 
\usepackage{mathtools}
\usepackage{float}

\usepackage{xcolor} 
\usepackage{amsmath,amsfonts,amssymb} 
\usepackage{siunitx}
\usepackage{tensor}
\usepackage{caption} 
\usepackage{subcaption}
\usepackage{dsfont}
\usepackage{braket}
\usepackage{dcolumn}
\usepackage{bm}
\DeclareMathAlphabet{\mathpzc}{OT1}{pzc}{m}{it}

\usepackage[bookmarks=true, pdfpagelabels=true, colorlinks=true, linkcolor=black, urlcolor=black, citecolor=blue, anchorcolor=blue]{hyperref}
\usepackage[capitalise,english]{cleveref}

\newcommand{\LL}{\mathcal{L}}
\newcommand{\OO}{\mathcal{O}}

\newcommand*{\software}[1]{\texttt{#1}}

\DeclareMathOperator{\artanh}{artanh}

\begin{document}

\preprint{CP3-22-14}

\title{Quantum SMEFT tomography: top quark pair production at the LHC}

\author{Rafael Aoude}
\email{rafael.aoude@uclouvain.be}
\affiliation{Centre for Cosmology, Particle Physics and Phenomenology (CP3),  Universit\'e catholique de Louvain, 1348 Louvain-la-Neuve, Belgium}

\author{Eric Madge}
\email{eric.madge-pimentel@weizmann.ac.il}
\affiliation{%
Department of Particle Physics and Astrophysics,
Weizmann Institute of Science, Rehovot 7610001, Israel%
}

\author{Fabio Maltoni}
\email{fabio.maltoni@uclouvain.be}
\affiliation{Centre for Cosmology, Particle Physics and Phenomenology (CP3),  Universit\'e catholique de Louvain, 1348 Louvain-la-Neuve, Belgium}
\affiliation{Dipartimento di Fisica e Astronomia, Universit\`a di Bologna and INFN,
Sezione di Bologna, via Irnerio 46, 40126 Bologna, Italy}

\author{Luca Mantani}
\email{luca.mantani@maths.cam.ac.uk}
\affiliation{
DAMTP, University of Cambridge, Wilberforce Road, Cambridge, CB3 0WA, United Kingdom}

\date{\today}

\begin{abstract}
Quantum information observables, such as entanglement measures, provide a powerful way to characterize the properties of quantum states. We propose to use them to probe the structure of fundamental interactions and to search for new physics at high energy. 
Inspired by recent proposals to measure  entanglement of top quark pairs produced at the LHC, we examine how higher-dimensional operators in the framework of the SMEFT modify the Standard Model expectations. We explore two regions of interest in the phase space where the Standard Model produces maximally entangled states: at threshold and in the high-energy limit. We unveil a non-trivial pattern of effects, which depend on the initial state partons, $q\bar q$ or $gg$, on whether only linear or up to quadratic SMEFT contributions are included, and on the phase space region. In general, we find that higher-dimensional effects lower the entanglement predicted in the Standard Model. 
\end{abstract}


\maketitle


\section{Introduction}
\label{sec:Introduction}

In 1989, Wheeler suggested that all physical quantities have a theoretical information origin, a concept which has been later popularized as the "\textit{it from bit}" principle~\cite{Wheeler1,Wheeler2}. Quantum Information~(QI) theory provides us with a set of tools and observables that are designed to unveil the inner behaviour of quantum mechanics. While these phenomena have been widely tested in applications at the atomic and even human scales, their study at higher energies has not been undertaken. 

Central to quantum information is entanglement, {\it i.e.}, the property of quantum systems to maintain a correlation even when they are separated.  Since the groundbreaking papers by Einstein, Podolski, and Rosen~\cite{PhysRevA.40.4277}, the series of papers by Schr\"odinger~\cite{Schrodinger:1935zz,schrodinger_1935,schrodinger_1936} and Bell's Theorem~\cite{PhysicsPhysiqueFizika.1.195}, entanglement has evolved from a puzzling and uncomfortable phenomenon to the keystone of quantum computation, as recently demonstrated by the quantum supremacy~\cite{Arute:2019}.

Vastly studied across the fields, ranging from wormholes~\cite{Maldacena:2013xja} to bacteria~\cite{Marletto_2018},
QI based analyses have led to significant advances not only in computation but also in fundamental questions, such as the black hole information paradox~\cite{PhysRevD.14.2460}. Many of such advances characterize new approaches in outstanding problems in string theory as well as in quantum field theory~\cite{Lykken:2020xtx}. In the latter case, for instance, a conjecture linking entanglement suppression and the presence of symmetries was proposed based on an observation for nuclear forces~\cite{PhysRevLett.122.102001} and later was also seen in the context of black hole physics~\cite{Aoude:2020mlg,Chen:2021huj} (see also Ref.~\cite{Low:2021ufv}). 
Proposals for understanding properties of the Standard Model~(SM) of particle physics based on QI properties have also appeared~\cite{Cervera-Lierta:2017tdt}. 

In this work we propose to exploit entanglement through specific measures, {\it e.g.}, the concurrence, to constrain new physics effects on the strength and type of interactions between SM particles in the context of the SM Effective Field Theory (SMEFT). As a first application of this idea, we consider the production of a top quark pair at the LHC and, in particular, the quantum state of their spins, which, as shown in Refs.~\cite{Afik:2020onf,Severi:2021cnj,Afik:2022kwm}, can be fully reconstructed from the decay products of the top and anti-top quarks and can be observed with high statistical significance already with Run II data. 
Further studies on the violation of Bell inequalities in top-quark pair production~\cite{Fabbrichesi:2021npl,Severi:2021cnj} and Higgs decays to $WW$~\cite{Barr:2021zcp} highlighted the potential of high-energy measurements in establishing the quantum nature of fundamental interactions. 

Contrary to the other quarks, the top quark decays way before its spin is affected by hadronization. The spin information is then imprinted in the decay products, and in particular in direction of the charged lepton from the $W$ decay, which is 100\% correlated with the corresponding top spin~\cite{Mahlon:2010gw}.
Another advantage of top quark pairs is that they can be characterized as simple bipartite qubit systems. 
Refs.~\cite{Afik:2020onf,Afik:2022kwm} outlined an experimental detection strategy for the concurrence of the top pair, based on measuring the differential cross-section with respect to the angular separation of the charged leptons from the top decays as a function of an upper cut on the top pair's invariant mass~\footnote{%
Note that, despite the incomplete knowledge of the momenta of the decay products in leptonic top decays, the top pair can be reconstructed with an efficiency of $\sim(80 - 90)\,\%$~\cite{ATLAS:2019zrq,CMS:2019nrx}.%
}.
As the same strategy can also be applied in the presence of new physics, we here limit ourselves to studying the influence of new physics on the entanglement, and neglect detector effects and other experimental issues.

In Ref.~\cite{Afik:2020onf}, it was further shown that there exist two regions of the $t\bar{t}$ phase-space where the top quark spins are maximally entangled: at threshold, when the partonic center-of-mass energy $\hat{s}=2m_t$, and for high energies and $\theta=\pi/2$, where the characteristic quantum states are Bell states, singlet and triplet, respectively.

With the aim of establishing how new physics might induce a modification of the quantum correlations, we study the production of a top quark pair system  within the SMEFT framework,  which provides a model independent formulation of new interactions when the energy scale associated to new physics is well separated from the scale of the process. In order to gain a detailed understanding, we compute the analytical linear and quadratic effects generated by dimension-six operators on the entanglement. We find it most useful to explore their behaviour at phase-space points where the entanglement in the SM is maximal. In doing so,  two immediate questions arise: \textit{Are these SM's maximally-entangled regions affected by the SMEFT?} \textit{Can SMEFT induce new maximally-entangled regions in the phase-space?}

The paper is organized as follows. In \cref{sec:TopSpinCorrelations}, we review the entanglement concepts and measures in bipartite systems which are then applied to top quark pair production later. In  \cref{sec:TopEntanglementSMEFT} we explore the effects of SMEFT higher-dimensional operators at linear and quadratic level in the full phase-space as well as for the angular-averaged concurrence. Finally, in \cref{sec:QuantumState} we explore the quantum state in different phase space regions. We draw our conclusions in \cref{sec:Conclusions}.

\section{Top pair spin correlations}
\label{sec:TopSpinCorrelations}

Central to the analysis of top-pair spin correlations is the spin production density matrix, also known as the $R$-matrix
\begin{align}
    \label{eq:Rmatrix_Amplitudes}
    R^{I}_{\eta_1\eta_2,\zeta_1\zeta_2} &\equiv \frac{1}{N_a N_b}\sum_{\substack{\text{colors}\\\text{a,b spins}}} \hspace{-.5em} \mathcal{M}_{\eta_2 \zeta_2}^*\,  \mathcal{M}_{\eta_1 \zeta_1}
\end{align}
with $\mathcal{M}_{\eta\zeta} \equiv \langle t(k_1,\eta)\bar{t}(k_2,\zeta)|\mathcal{T}|a(p_1)b(p_2)\rangle$,
where $\mathcal{T}$ is the transition matrix element, $I=ab$ denotes the initial state, $N_{a,b}$ is the number of degrees of freedom of the respective initial state particles $a$ and $b$, $k_i$ ($p_i$) are the momenta of the final (initial) state particles, and $\eta$ ($\zeta$) are the (anti-)top spin indices.
Note that this matrix is similar to the cross-section, but with uncontracted final-state spin-indices. 
The $R$-matrix is commonly expressed in terms of the Pauli matrices $\sigma^i$, also known as the Fano decomposition~\cite{RevModPhys.55.855}, which reads
\begin{equation}
\label{eq:FanoDecomposition}
R = \tilde{A}\, \mathds{1}_2\otimes \mathds{1}_2
+ \tilde{B}_i^+\sigma^i\otimes \mathds{1}_2 
+ \tilde{B}_i^- \mathds{1}_2 \otimes \sigma^i
+ \tilde{C}_{ij}\,\sigma^i\otimes \sigma^j ,
\end{equation}
where a sum over $i,j = 1,2,3$ is implicit.
The spin-correlations between the sub-systems are captured by the $\tilde{C}_{i,j}$ coefficients, while the $\tilde{B}_i^\pm$ coefficients describe the degree of (anti-)top polarization, and $\tilde{A}$ is related to the differential cross-section by
\begin{equation}
    \frac{\mathrm{d} \sigma}{\mathrm{d} \Omega \mathrm{d} \hat{s}}=\frac{\alpha_{s}^{2} \beta}{\hat{s}^{2}} \tilde{A}\left(\hat{s}, \bm{k}\right) \, ,
\end{equation}
where $\bm{k}$ is the top quark direction, $\hat{s}$ the invariant mass of the top quark pair and $\beta = \sqrt{1-4m_t^2/\hat{s}}$ the velocity of the top quark in the center-of-mass frame with the top mass $m_t=\SI{172.76}{\GeV}$~\cite{ParticleDataGroup:2020ssz}. Altogether, there are $15$ Fano coefficients which, once determined, allow us to fully characterize the quantum state of the two-qubit system.

In the following, we consider top quark pair production at the LHC in proton proton collisions, where the top pairs are at leading order~(LO) created through non-interfering channels.
The relevant contributing initial states are the scattering of two gluons and of a quark anti-quark pair.
As a result, the quantum state of the system is mixed, with the total density matrix given by the weighted sum of the channel-specific matrices. 
The weights depend on the structure of the proton described by the parton distribution functions.
The full $R$-matrix is hence given by the sum over the partonic channels, each weighted by the corresponding luminosity functions~$L^I(\hat{s})$
\begin{align}
R(\hat{s}, \bm{k}) = \sum_I L^I(\hat{s}) R^I(\hat{s},\bm{k}) \,.
\label{eq:RmatrixWeightedLuminosity}
\end{align}
The $gg$-initiated channel dominates up to \si{TeV} top quark energies, when the $q\bar{q}$ channel becomes comparable.

The density matrix describing the quantum state is given by the normalized $R$-matrix, {\it i.e.}, $\rho=R/\text{tr}(R)$, which we expand in terms of the coefficients $B_i^\pm=\tilde{B}_i^\pm/\tilde{A}$ and $C_{ij} = \tilde{C}_{ij}/\tilde{A}$ as
\begin{equation}
\label{eq:rho}
\rho = \frac{\mathds{1}_2 \!\otimes\! \mathds{1}_2
+ B_i^+\sigma^i \!\otimes\! \mathds{1}_2 
+ B_i^- \mathds{1}_2 \!\otimes\! \sigma^i
+ C_{ij}\,\sigma^i \!\otimes\! \sigma^j}{4} .
\end{equation}

In order to obtain explicit values for the entanglement, we calculate the coefficients in the so-called helicity basis, which consists of an orthonormal basis in the centre-of-mass frame
\begin{align}
\label{eq:helicity_basis}
\{\bm{k},\bm{n},\bm{r}\}: 
\,\,
\bm{r} = \frac{(\bm{p} - z\bm{k})}{\sqrt{1-z^2}},
\quad
\bm{n} = \bm{k}\times \bm{r},
\end{align}
where $\bm{p}$ and $\bm{k}$ are the unit vectors along the beam axis and top quark direction, and we define $z \equiv \bm{k}\cdot \bm{p} =\cos\theta$.
For convenience, we switch to the variables $z$ and $\beta$ for the remainder of the presentation.
In this basis, the spin density matrix for top quark pair production in the SM at LO in QCD simplifies significantly~\cite{Bernreuther:2015yna}: invariance under $CP$ renders $C_{ij}$ symmetric and $B_i^+=B_i^-$, and non-zero $C_{kn}$, $C_{rn}$ and $B^\pm_n$ are then only generated at the one-loop level by absorptive parts.
Furthermore, $B_k^\pm$ and $B_r^\pm$ vanish as they require $P$-odd interactions.
As we will focus only on $CP$ even operators, the first two statements also hold true when adding the SMEFT contributions.

\subsection{Entanglement}
\label{sec:Bipartite}

The most general bipartite quantum state is described by a normalized density matrix $\rho \in \mathcal{D}(\mathcal{H}_{\rm ab})$, where the space $\mathcal{D}(\mathcal{H}_{\rm ab})$ is formed by the positive-semidefinite operators acting on the full system's Hilbert space ${\cal H}_{\rm ab} ={\cal H}_{\rm a}\otimes {\cal H}_{\rm b}$. Whenever the state can be written as a convex combination of product states~\cite{PhysRevA.40.4277}
\begin{align}
\rho_{\rm ab}  = \sum_{k} p_k\, \rho^k_{\rm a}\otimes \rho^k_{\rm b}
\end{align}
the state is said to be separable. A state which cannot be written is such form is said to be \textit{entangled}. This notion was introduced by Schr\"odinger~\cite{schrodinger_1935} under the name of Verschr\"ankung to describe that the best knowledge of the whole system ${\cal H}_{\rm ab}$ does not imply the best possible knowledge of its parts, ${\cal H}_{\rm a}$ and ${\cal H}_{\rm b}$. These formal definitions are more transparent when a quantitative measure of entanglement is given.

Several entanglement measures and criteria are available, cf.\ e.g.\ Refs.~\cite{bengtsson_zyczkowski_2006,Horodecki:2009zz} for a detailed discussion. In the following, we quantify the degree of entanglement by defining a physical quantity called concurrence~\cite{Wootters:1997id}
\begin{equation}
	C[\rho] = \max\left(0,\lambda_1-\lambda_2-\lambda_3-\lambda_4\right)\,,
\end{equation}
where $\lambda_i$ are the increasingly ordered eigenvalues of the matrix $\omega = \sqrt{\sqrt{\tilde{\rho}}\rho\sqrt{\tilde{\rho}}}$ with $\tilde{\rho} = (\sigma_2\otimes\sigma_2)\rho^*(\sigma_2\otimes\sigma_2)$ and $\rho^*$ denoting the complex conjugate of $\rho$, or, equivalently, the square roots of the eigenvalues of the matrix $\rho \tilde{\rho}$. When $C[\rho] > 0$, the system is said to be entangled and the case of $C[\rho]=1$ corresponds to quantum configurations of maximal entanglement.

As a simpler criterion for entanglement, we further employ the Peres-Horodecki Criterion~(PHC)~\cite{Peres:1996dw,Horodecki:1997vt}, which states that $\rho$ is entangled, if the partial-transpose state
\begin{align}
\rho_{\mathrm{sep}}^{\rm T_a} \equiv ({\rm T} \otimes \mathds{1})(\rho_{\mathrm{sep}})=
\sum_{k} p_k\, (\rho^k_{\rm a})^{\rm T} \otimes \rho^k_{\rm b} \geq 0
\end{align}
is non-negative with unit trace. 
This is a necessary condition for entanglement for $2\otimes 2$ bipartite systems.

For the density matrix in \cref{eq:rho}, in the helicity basis, the PHC implies~\cite{Afik:2020onf}
\begin{align}
\label{eq:DeltaPeresHorodecki}
\Delta \equiv  - C_{nn} + |C_{kk}  + C_{rr}| -1 > 0
\end{align}
as a sufficient condition for entanglement. 
For the SM at LO in QCD, $\Delta > 0$ is a necessary condition and the concurrence can be written in terms of $\Delta$ as $C[\rho] = \max(\Delta/2,0)$.
The corresponding expression including SMEFT corrections is obtained by expanding $\omega$ in $1/\Lambda^2$ to quadratic order, where $\Lambda$ is the new physics scale, and can be found in \cref{app:Concurrence}.

The correlation matrix further simplifies when averaging over the solid angle. 
Switching to the beam basis and defining the angular averaged $R$-matrix following Ref.~\cite{Afik:2020onf},
\begin{align}
\label{eq:AverageRmatrix} 
    \bar{R} = (4\pi)^{-1}\int {\rm d}\Omega\,R(\hat{s} ,\bm{k})\,,
\end{align}
as well as the corresponding density matrix $\bar{\rho}=\bar{R}/\mathrm{tr}(\bar{R})$, the correlation matrix becomes diagonal with two degenerate eigenvalues, $C_{ij} = \mathrm{diag}(C_\perp, C_\perp, C_{z})$, and the only non-vanishing entry in $B^\pm_i$ is in the $z$ component. 
The PHC then implies the sufficient condition~\cite{Afik:2020onf}
\begin{align}
\label{eq:deltaPeresHorodecki}
\delta \equiv -C_z + |2C_{\perp}| -1 > 0
\end{align}
for entanglement, which becomes a necessary condition for $B^\pm_z = 0$ with the concurrence given by $C[\rho] = \max(\delta/2,0)$.
The expression for the concurrence in the SMEFT with $B^\pm_z \neq 0$ is presented in \cref{app:Concurrence}.

\subsection{Quantum states}
\label{sec:QuantumStateSM}

\begin{figure}
	\includegraphics[width=0.95\linewidth]{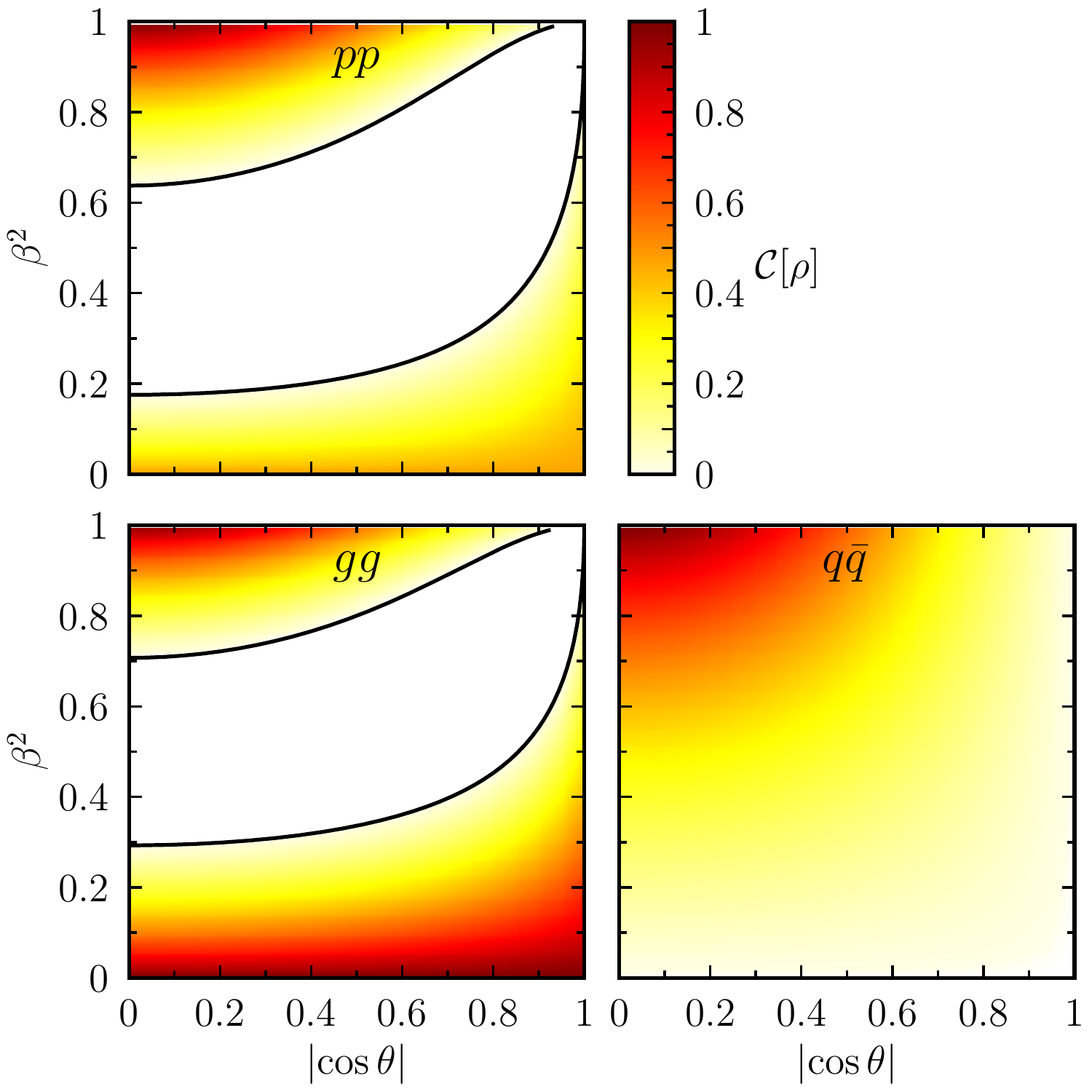}\,
	\caption{SM contribution for the concurrence in the gluon-~(bottom left) and quark-initiated~(bottom right) channels, as well as in the full $pp$ collision~(top). The black lines indicate the boundaries of the entangled regions based on \cref{eq:DeltaPeresHorodecki}.
	}
	\label{fig:Concurrence_SM}
\end{figure}

The SM contribution to the concurrence in terms of $\beta^2$ and $z=\cos\theta$ is shown in Fig.~\ref{fig:Concurrence_SM}.
In the $gg$-channel, the top quark pair is produced in a maximally entangled state at production threshold, {\it i.e.}, for $\beta=0$ at any value of $z$, as well as at high transverse momentum, {\it i.e.}, when $\beta=1$ and $\cos\theta=0$. As noted in~\cite{Afik:2020onf}, the gluon-induced quantum state $\rho^{\rm SM}_{gg}(\beta,z)$ becomes a Bell state in these regions: at threshold it corresponds to a singlet state, while at high energy it is a triplet state along the $\bm{n}$-axis
\begin{align}
\label{eq:QuantumStateSM}
    \rho_{gg}^{\text{SM}}(0,z) = |\Psi^-\rangle_{\bm n} \langle \Psi^-|_{\bm n},\quad
    \rho_{gg}^{\text{SM}}(1,0) = |\Psi^+\rangle_{\bm n} \langle \Psi^+|_{\bm n},
\end{align}
where the density matrices are defined in terms of the Bell basis (or Einstein–Podolsky–Rosen pairs) along the $\bm{n}$-axis
\begin{align}
    |\Phi^\pm \rangle_{\bm n} = \frac{|\!\uparrow \uparrow\rangle_{\bm{n}} \!\pm
    |\!\downarrow\downarrow\rangle_{\bm{n}} }{\sqrt{2}},
    \quad
    |\Psi^\pm \rangle_{\bm n} = \frac{|\!\uparrow \downarrow\rangle_{\bm{n}}  \!\pm|\!\downarrow \uparrow \rangle_{\bm{n}}}{\sqrt{2}}
\end{align}
and similarly along the $\bm{p}$-axis.
In the $q\bar q$ initiated channel, the top-pair is entangled across all phase-space except at threshold, where we have a mixed but separable state. At high energies, the same maximally entangled state as in the $gg$-channel is produced
\begin{align} 
\rho_{q\bar q}^{\text{SM}}(1,0) = |\Psi^+\rangle_{\bm n} \langle \Psi^+|_{\bm n}.
\end{align}

Simulation-based analyses of the top-pair entanglement in the SM have been performed in  Refs.~\cite{Afik:2020onf,Severi:2021cnj}, where evidence was shown for such effects to be detected at the LHC with high statistical significance. We assume that the same strategies used to reconstruct the quantum states from the decay products of the top quarks  remain valid when adding SMEFT operators, since these operators do not affect the correlations between the top spin and the charged lepton decay product~\cite{Zhang:2010dr}.

\section{Entanglement in the SMEFT}
\label{sec:TopEntanglementSMEFT}

To study the impact of higher-dimensional operators on the entanglement in top quark pair production within the SMEFT, we use a slightly modified version of the Warsaw basis~\cite{Grzadkowski:2010es}, detailed in~\cite{Degrande:2020evl}, 
\begin{align}
\LL_{\rm SMEFT} = \LL_{\rm SM} +  \frac{1}{\Lambda^2}\sum_i c_i \OO_i \,,
\end{align}
where we restrict ourselves to $CP$-even operators at dimension-six.
Working at LO in QCD, the relevant operators to $t\bar{t}$ production are the zero- and two-fermion operators
$\OO_{G}, \OO_{\varphi G}$ and $\OO_{tG}$, as well as the two-light-two-heavy four-fermion octets and singlets~\cite{Zhang:2010dr,Degrande:2010kt,Degrande:2020evl,Bernreuther:2015yna}, whose definitions are collected in~\cref{app:SpinDensity}. 
In order to have a tractable amount of four-fermion operators, we have imposed a $U(2)_q \otimes U(2)_u \otimes U(3)_d$ flavour symmetry~\cite{Aguilar-Saavedra:2018ksv}.
For the purpose of illustration, we chose values of the Wilson coefficients that may exceed the current limits from the \software{SMEFit} collaboration~\cite{Ethier:2021bye} in most of our plots. The conclusions however remain true also for values within these bounds.

We now turn to investigating the effects of dimension-six SMEFT operators on the entanglement. 
Such operators will lead to an EFT contribution to the $R$-matrix as well as the density matrix,
\begin{align}
\label{eq:rhoSMEFT}
\rho = \frac{R^{\rm SM}+ R^{\rm EFT}}{\text{tr}(R^{\rm SM})+\text{tr}(R^{\rm EFT})}\,.
\end{align}
The contributions of each SMEFT operator considered here to the Fano coefficients are given in \cref{app:SpinDensity}. 
The concurrence and the entanglement markers in \cref{eq:DeltaPeresHorodecki,eq:deltaPeresHorodecki} are then obtained expanding in the Wilson coefficients.
The corresponding expressions can be found in \cref{app:Concurrence}.
In the following, we will study the entanglement taking into account both linear, $\OO (\Lambda^{-2})$, and quadratic effects%
~\footnote{The latter are strictly speaking only a subset of all $\OO (\Lambda^{-4})$ contributions, which in addition also include the double-insertions and dimension-eight operators not considered here.}%
, $\OO (\Lambda^{-4})$, to these quantities. 
Then, averaging over the solid angle, we further present the entanglement as a function solely of the top-quark velocity~$\beta$. 

\subsection{Linear interference and quadratic effects}

To examine the impact of the linear and quadratic SMEFT corrections on the entanglement, we consider the PHC and the entanglement marker~\cref{eq:DeltaPeresHorodecki} in SMEFT and compare it to the marker $\Delta_0$ in the SM. 
Since the absolute value in \cref{eq:DeltaPeresHorodecki} does not allow to factor out the Wilson coefficients, we define at linear order $\Delta_1 \equiv \Delta - \Delta_0$, where $\Delta$ is calculated from the density matrix~\cref{eq:rhoSMEFT} including the SM and linear corrections.  
Equivalently, at the quadratic order, we define $\Delta_2 \equiv \Delta - (\Delta_0 + \Delta_1)$, where we now also include the dimension-six squared contributions to $\rho$ in $\Delta$.

\begin{figure}
	\includegraphics[width=.98\linewidth]{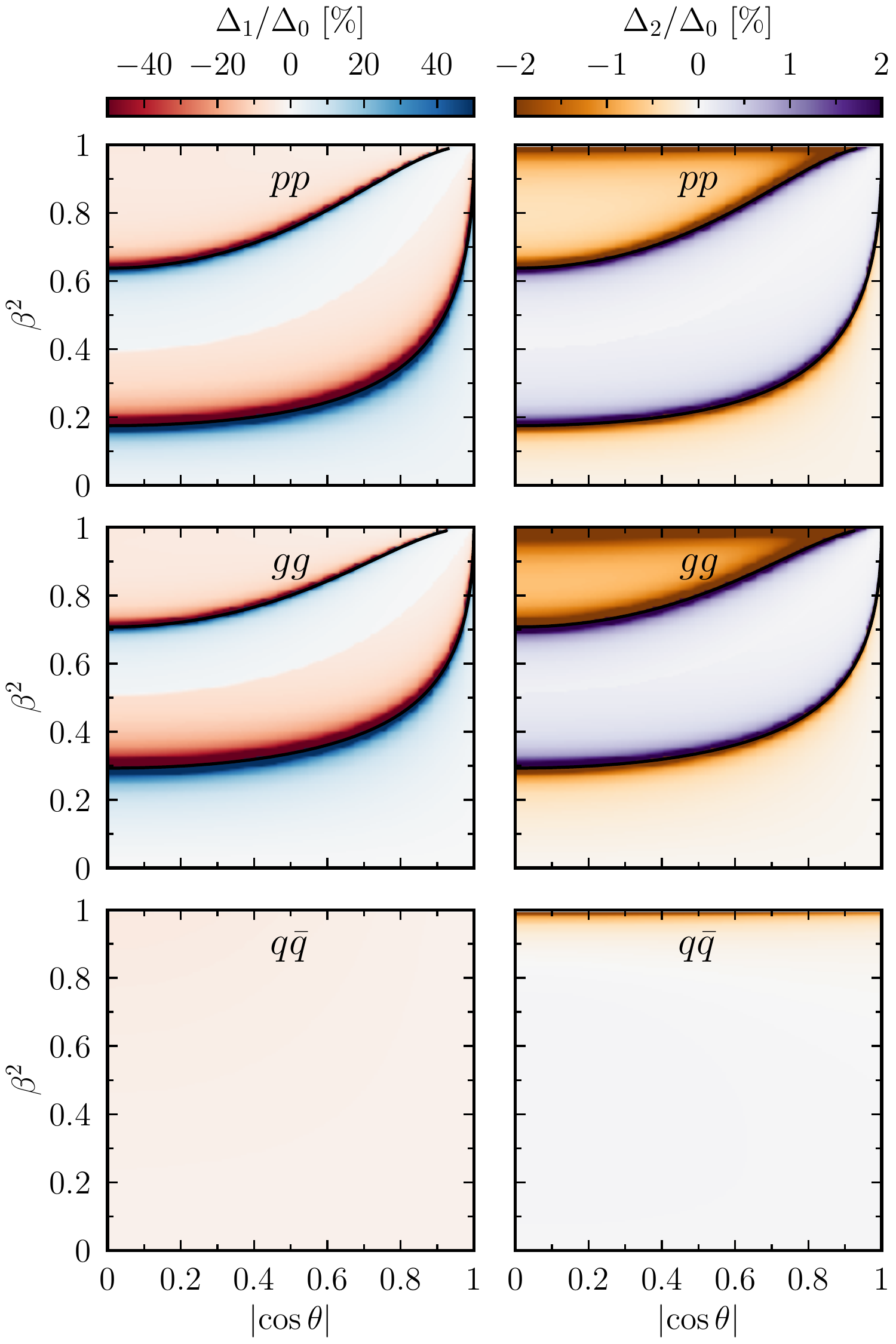}\,
	\caption{Relative contributions of the linear~(left) and quadratic~(right) effects of the chromomagnetic operator for $c_{tG}/\Lambda^2   =\SI{0.1}{\TeV^{-2}}$ to the PHC entanglement marker $\Delta$.}
	\label{fig:ctG_Ratios}
\end{figure}

\Cref{fig:ctG_Ratios} depicts the new physics contributions of the operator $\OO_{tG}$ to $\Delta$ relative to the SM value for $c_{tG}=\SI{0.1}{\TeV^{-2}}$, which lies within the limits from current fits~\cite{Ethier:2021bye}.
Note that $\Delta$ becomes negative in the absence of entanglement, whereas the concurrence vanishes in the entire unentangled region. 
Hence, we can still take the ratios $\Delta_{1,2}/\Delta_0$ in this region.
However, at the boundary of the entangled phase space (along the black lines in \cref{fig:ctG_Ratios}), $\Delta$ vanishes and the ratios diverge.

Let us start our discussion with the $gg$-channel, where only $\OO_{tG}$, $\OO_{G}$ and $\OO_{\varphi G}$ contribute. 
At leading order, we notice that the SM's point of maximal entanglement at threshold is unchanged by linear interference effects, as can be seen in the middle left panel of \cref{fig:ctG_Ratios}, where the concurrence is zero for $\beta \rightarrow 0$.
This is also apparent from the $\beta$ dependence of the Fano coefficients, explicitly given in \cref{app:SpinDensitySMEFT_Lambda2}. 
As both $c_{G}$ and $c_{\varphi G}$ always come accompanied by a factor of $\beta^2$, it is evident that their contributions vanish at threshold to order $\OO(\Lambda^{-2})$. 
While the contributions of $\OO_{tG}$ to the individual Fano coefficients survive for $\beta \rightarrow 0$, they cancel at the concurrence level, giving zero contribution to the entanglement. 
Hence, including linear effects, top quark pairs produced in the $gg$-channel remain maximally entangled at threshold.

At quadratic order $\OO(\Lambda^{-4})$, on the other hand, a different behaviour emerges.  
Although the contribution from $\OO_{\varphi G}$ vanishes at threshold, the other operators $\OO_{G}$ and $\OO_{tG}$ decrease the concurrence at the point of maximal entanglement of the SM and induce a triplet state on top of the SM's singlet state. This effect will be discussed in further detail in \cref{sec:QuantumState}.

The other point of maximal entanglement in the SM, $\beta \rightarrow 1$ and $z=0$, 
where the SM produces a triplet state
, is more delicate.
As $\beta\to1$ corresponds to $\hat{s} \rightarrow \infty$, the EFT is not valid at this point.
We can however consider a region where $\beta \sim 1$ but $m_t^2 \ll \hat{s} \ll \Lambda^2$, such that  top quarks can be considered massless, yet within the regime of validity of the EFT. 
Then, for all of the zero- and two-fermion operators we consider, the amount of entanglement produced in the SM is decreased by the squared SMEFT correction, while the behaviour of the linear interference terms depends on the sign of the corresponding Wilson coefficient. 
For sufficiently large negative values of the coefficient, this point of maximal entanglement can extend to a larger region.

Moving on to the $q\bar{q}$-initiated channel, contributions arise from $\OO_{tG}$ as well as from the four-fermion operators.
The latter can be classified based on their behaviour upon top-quark charge conjugation, see \cref{app:SpinDensitySMEFT_Lambda2} and Ref.~\cite{Brivio:2019ius}.
At linear order, we observe that $\tilde{A}_{\rm EFT}^{q\bar{q}}$ and $\tilde{C}_{ij,\rm{EFT}}^{q\bar{q}}$ are only affected by the combination of operators corresponding to two vector or two axial-vector currents, whereas the combinations consisting of one vector and one axial-vector current only enter in the $B^{\pm}_i$ coefficients. The SMEFT corrections at quadratic order are more involved, and in particular also include contributions from the singlet operators. 
The analytical results can be found in~\cref{app:SpinDensitySMEFT_Lambda4}. 

The linear and squared corrections from the SMEFT in the $q\bar{q}$ channel are shown for the example of $\OO_{tG}$ in the lower panel of~\cref{fig:ctG_Ratios}.
There are no contributions to $\Delta$ at threshold, even at the quadratic level, whereas at high energies, the linear and quadratic effects may modify the level of entanglement around the SM point of maximal entanglement, where the contribution of the latter always decreases the concurrence at high $p_T$ (the former of course depend on the sign of the respective Wilson coefficient).

In order to explore these effects in detail, in the following section we study the angular-averaged concurrence and the explicit quantum states.

\subsection{Angular averaged concurrence}

To study the threshold region, we switch from the helicity basis \cref{eq:helicity_basis} to the beam basis (with the $z$ axis along the beam direction), and perform the angular integration as suggested in~\cite{Afik:2020onf}.
The analytical expressions for the Fano coefficients derived from the averaged density matrix in \cref{eq:AverageRmatrix} can be found in \cref{app:SpinDensityOmega}. 

\begin{figure}
 	\includegraphics[width=0.95\linewidth]{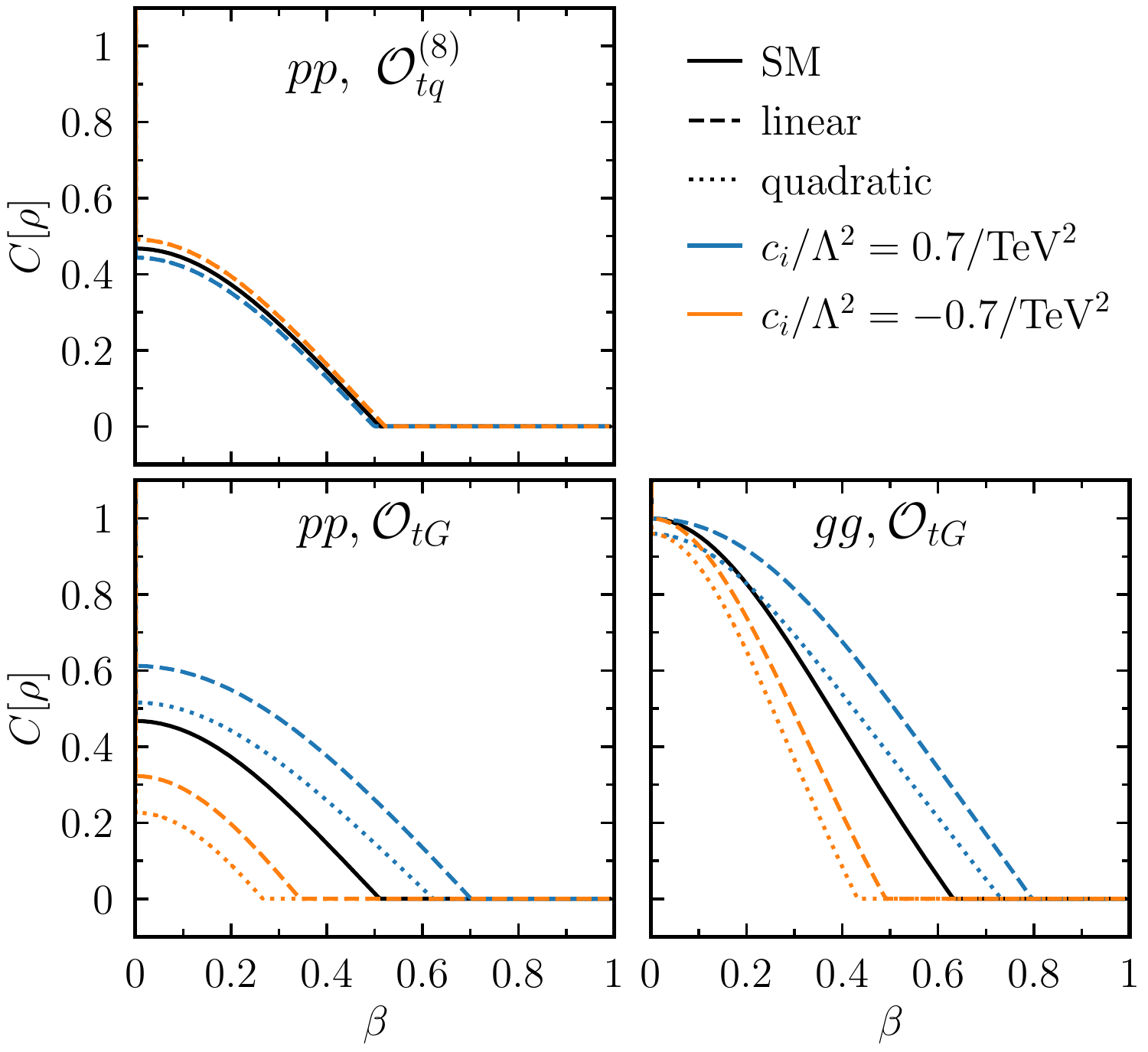}
 	\caption{Concurrence averaged over solid angle as a function of the top quark velocity $\beta$ for $c_{tq}^{(8)}$~(top) and $c_{tG}$~(bottom).}
 	\label{fig:Omega_1d}
 \end{figure}

The new physics effects on the corresponding concurrence as a function of $\beta$  are depicted in \cref{fig:Omega_1d} on the example of $\OO_{tG}$~(bottom) and $\OO_{tq}^{(8)}$~(top), setting the Wilson coefficients to $c_i/\Lambda^2 = \pm \SI{0.7}{\TeV^{-2}}$, where the blue~(orange) lines correspond to the positive~(negative) sign.
The black solid line indicates the value in the SM, whereas the dashed and dotted lines show the concurrence including linear and quadratic contributions.

Let us first consider the effects of $\OO_{tG}$ on $gg$-initiated production. 
In the SM, the angular-averaged concurrence decreases from $C[\rho]=1$ at threshold towards high energies as a function of $\beta$.
This behaviour is not modified by the EFT. 
However, the critical point where the entanglement marker crosses zero changes depending on the value of the Wilson coefficient due to the linear contribution, and is further modified by the squared corrections.
Furthermore, as already seen in \cref{fig:ctG_Ratios}, the linear EFT contributions do not affect at all the concurrence at threshold.
However, the effects from the squared operators $\OO_{tG}$ and $\OO_G$ can lead to a decrease of the level of entanglement, as can be seen in the lower right panel of~\cref{fig:Omega_1d}.
We delay a detailed discussion to this effect to~\cref{sec:QuantumState}. The operator $\OO_{\varphi G}$ does not induce any of such changes. 

In the $q\bar{q}$ channel, where the concurrence of the angular-averaged state vanishes in the SM, the corrections induced by the dimension-six operators are too small to induce entanglement. 
The new physics contribution in this channel may, however, still affect the concurrence when considering the total spin density matrix \cref{eq:RmatrixWeightedLuminosity} in $pp$ collisions. 
The effect here comes mostly from the corrections to $\tilde{A}$, as these determine the balance between the $gg$ and $q\bar{q}$ contributions to the total density matrix. 
Hence, in the case of $pp$ collisions displayed in the left panel of \cref{fig:Omega_1d}, we can observe an effect of the EFT already at threshold, even at the linear level.
This also holds true for four-fermion operators, such as for instance $\OO_{tq}^{(8)}$ shown in the upper panel.

\section{Quantum states in the SMEFT}
\label{sec:QuantumState}

In this section we consider the effects of new physics on the quantum state of the $t\bar{t}$ pair in different regions of the phase space. Observables directly related  to the quantum state probe different and complementary directions in the parameter space compared to the scattering amplitude. In the following, we discuss in particular two phase space regions of interest, {\it i.e.}, the production of top quarks at threshold, characterised by high statistics, and the production at high $p_T$ in the central region, where top-quark mass effects become negligible. 

\subsection{Threshold region}

In the SM, the $gg$ initiated channel at threshold is characterised by a pure maximally-entangled state as in~\cref{eq:QuantumStateSM}, with the top quark spins forming a singlet state of spin $0$. 
The presence of new physics effects can potentially change the picture. 
In particular, we find that the chromo-magnetic operator $\OO_{tG}$ and the triple-gluon operator~$\OO_G$ change the quantum state, which is then not a pure state anymore. As a matter of fact, these operators induce the presence of a triplet state of spin $1$, and the density matrix is therefore described by a mixed state:
\begin{align}
\rho_{gg}^{\rm EFT}(0,z) = p_{gg} |\Psi^+\rangle_{\bm p}\langle \Psi^+|_{\bm p}
+
(1-p_{gg})|\Psi^-\rangle_{\bm p}\langle \Psi^-|_{\bm p} \,.
\end{align}
Note that here the spins are defined with respect to the beam direction $\bm{p}$. 
The probability of being in a triplet state is given by
$p_{gg} = 72 m_t^2 (3\sqrt{2} m_t \, c_G + v \, c_{tG})^2/7\Lambda^4 \, ,$
which shows that no linear effects are present and only the squares contribute. 
In particular, we find a flat direction for a specific combination of $c_G$ and $c_{tG}$, while the operator $\mathcal{O}_{\varphi G}$ does not affect the quantum state at threshold.

For the $q\bar q$ channel, in the SM the spin density matrix is characterised by a mixed separable state:
\begin{equation}
    \rho_{q\bar{q}}^{\rm SM}(0,z) = \frac{1}{2}\,\ket{\uparrow \uparrow}_{\bm{p}}\bra{\uparrow \uparrow}_{\bm{p}} + \frac{1}{2}\,\ket{\downarrow \downarrow}_{\bm{p}}\bra{\downarrow \downarrow}_{\bm{p}} \, .
\end{equation}
Specifically, the probability of having both, top and anti-top quark, with spin up (down) is $1/2$ in the SM. The EFT effects in this case do not change the structure of the state, but the eigenvalues of the density matrix are affected and a preference for one spin direction is in general observed:
\begin{equation}
    \rho_{q\bar{q}}^{\rm EFT}(0,z) = p_{q \bar q} \ket{\uparrow \uparrow}_{\bm{p}}\bra{\uparrow \uparrow}_{\bm{p}} + (1-p_{q \bar q})\ket{\downarrow \downarrow}_{\bm{p}}\bra{\downarrow \downarrow}_{\bm{p}} \, ,
\end{equation}
where $p_{q\bar q} = \frac{1}{2}-4\frac{c_{VA}^{(8),u}}{\Lambda^2} + \OO(1/\Lambda^4)$, which also includes corrections at linear order in the Wilson coefficients~\footnote{The full expression for $p_{q\bar q}$ including the $\Lambda^4$ corrections can be found in~\cref{app:Probabilities}.}.
Here, $c_{VA}^{(8),u} = (-c_{Qq}^{(8,1)} -c_{Qq}^{(8,3)} + c_{tu}^{(8)} - c_{tq}^{(8)} + c_{Qu}^{(8)})/4$.
The spoiling of the symmetry is due to P-violating interactions induced by dimension-six operators but is also present if electroweak corrections are taken into account.

\begin{figure}
 	\centering
 	\includegraphics[width=.99\linewidth]{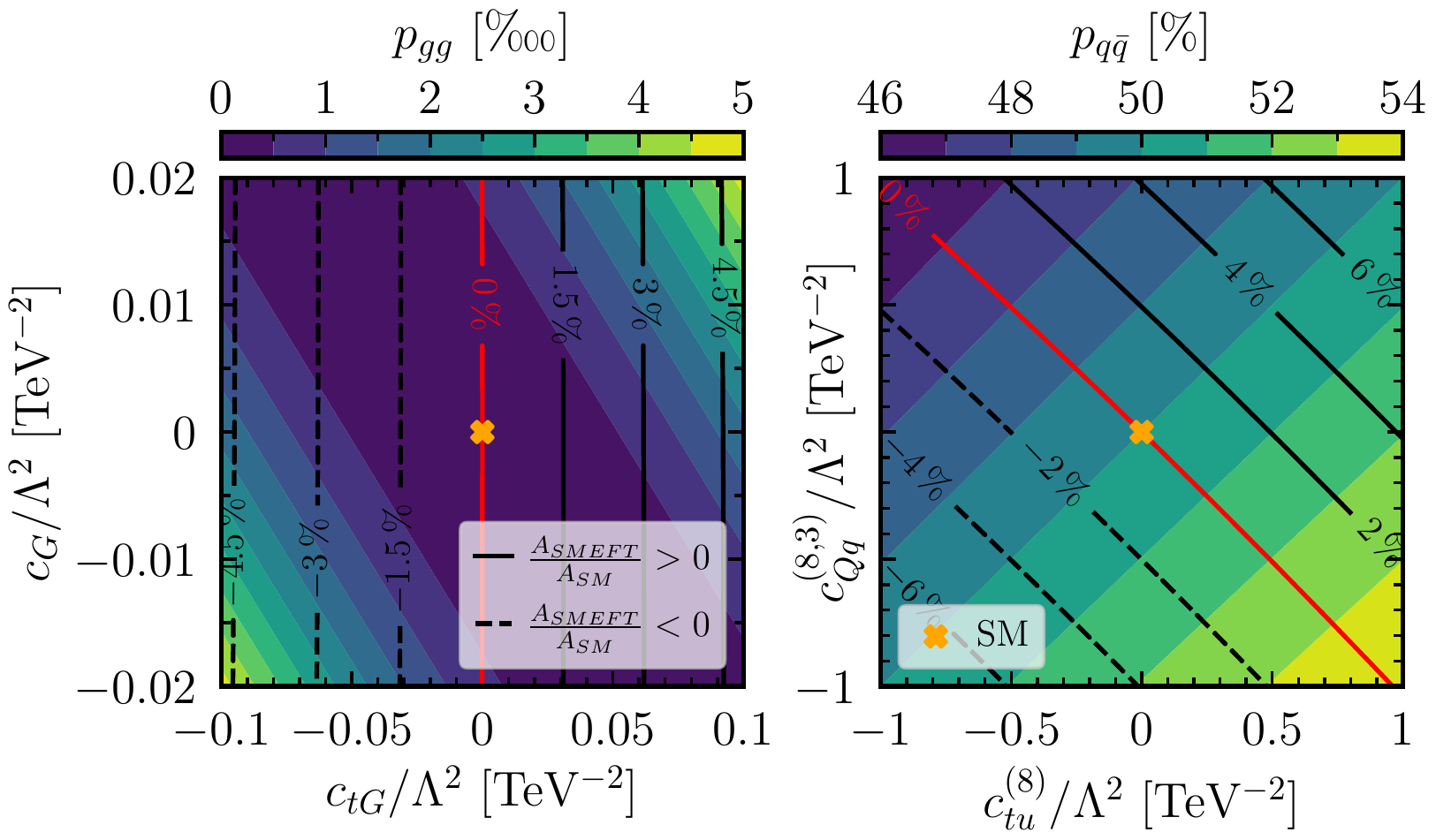}
 	\caption{Probability to produce a triplet~(left) or both-spin-up~(right) state at threshold ($\beta=0$) in the $gg$ or $q\bar{q}$ channel, respectively. The contour lines indicate the relative corrections of the EFT to the scattering amplitude.}
    \label{fig:prob_threshold}
 \end{figure}

In \cref{fig:prob_threshold} we show contour plots of the probabilities $p_{gg}$ and $p_{q\bar q}$. In the case of the quark initiated channel, we choose $\OO_{tu}^{(8)}$ and $\OO_{Qq}^{(8,3)}$ as a pair of representative four-fermion operators. In addition to the probabilities, we also plot contours of the relative EFT effects on the scattering amplitude, in order to highlight the complementarity of the two observables, which are clearly probing different directions in the parameter space.

%

\subsection{Central high-$p_T$ region}

The other interesting region to consider is the one characterised by high $p_T$. In the following, we set $\theta = \pi/2$ ($z=0)$ and look at the probability for the top quark pair to be in a triplet state
\begin{align}
p_{\Psi^+} = \langle \Psi^+|_{\bm n} \,  \rho \, |\Psi^+\rangle_{\bm n} \,,
\end{align}
which, in particular, is the quantum state for the SM in the limit of $\beta\to1$, both in the $gg$ and the $q \bar q$ initiated channels.
However, as already discussed above, the limit $\beta\to1$ is ill-defined in the presence of higher dimensional operators. 
We therefore study the probability as a function of the invariant mass $\hat{s}$ of the top quark pair (or partonic center-of-mass energy), where the EFT can be considered valid if $\hat{s} < \Lambda^2$.

\begin{figure}
    \centering
    \includegraphics[width=0.8\linewidth]{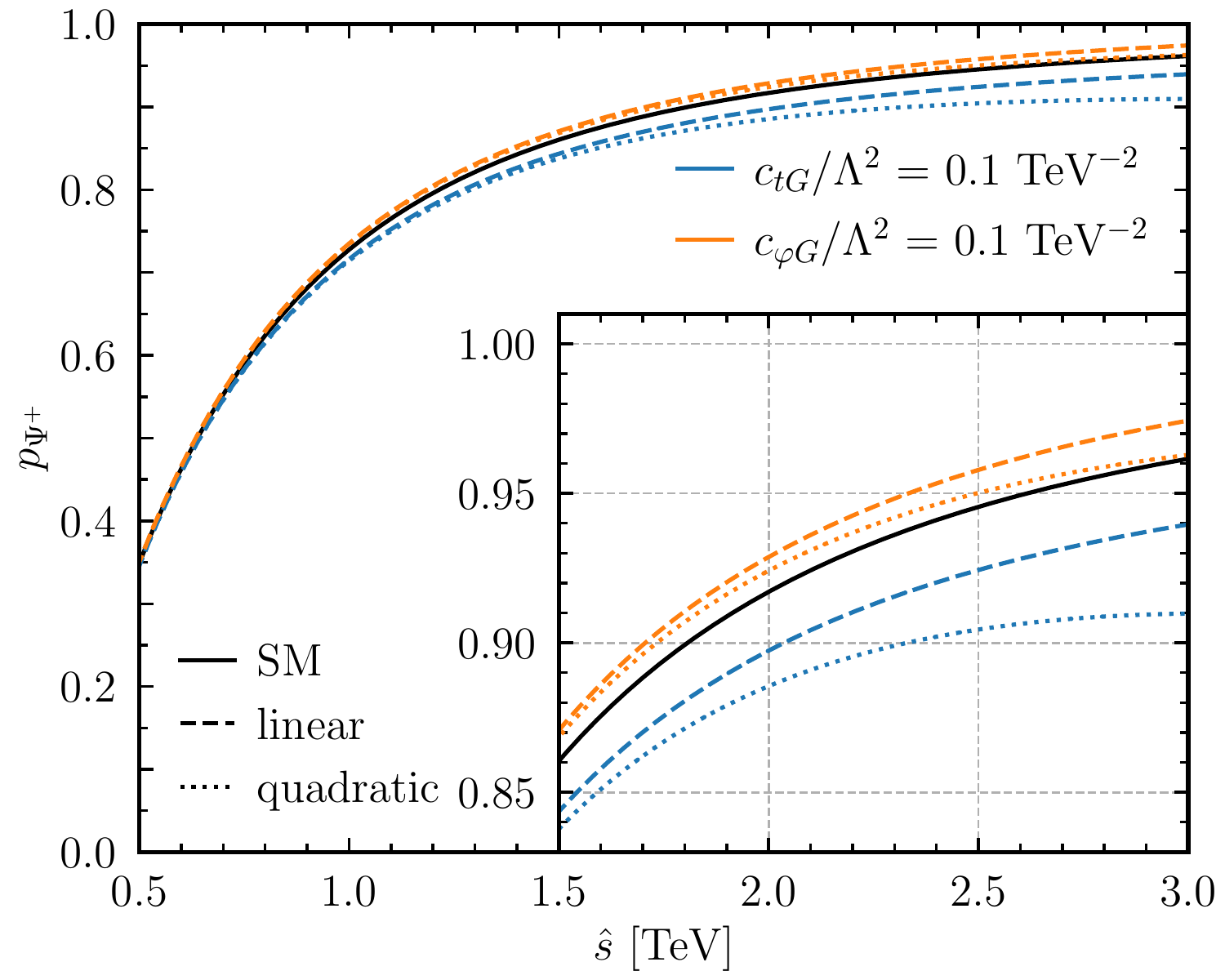}
    \caption{Probability for a triplet configuration in the $gg$ channel at $\cos\theta=0$, including linear and quadratic SMEFT contributions. The inlay zooms in on high energies.}
    \label{fig:central_triplet_gg}
\end{figure}

In \cref{fig:central_triplet_gg}, we plot the probability of the quantum state to be in a triplet configuration for the $gg$ channel, depicting the linear~(dashed) and quadratic~(dotted) effects of $\OO_{tG}$~(blue) and $\OO_{\phi G}$~(orange) for a Wilson coefficient of $c_i/\Lambda^2 = \SI{0.1}{\TeV^{-2}}$.
As we can observe, the probability in the SM converges towards~$1$, while the EFT effects become particularly manifest at high energy where we start to see deviations in the probability. Similar effects are observed in the quark/anti-quark initiated channel. Once again, the interesting aspect is that the observables related to the quantum state are probing a different direction in the parameter space with respect to the scattering amplitude, offering new and promising ways to disentangle the various higher-dimensional effects.

\section{Conclusions}
\label{sec:Conclusions}

We have proposed to use quantum observables, such as entanglement, to assess the impact of new physics effects in high-energy interactions. Using two measures, we have explored how linear and quadratic effects from SMEFT dimension-six operators affect entanglement in top quark pairs at the LHC. In particular, we have focused on two phase-space regions, at threshold and at high energies, where the SM produces maximally entangled states.

With entanglement being at the core of quantum mechanics, one might hope that it will provide fundamental information on the structure of the effective field theory as much as unitarity, analyticity and positivity do on general properties of the scattering amplitudes. At a more practical level quantum observables probe different directions in the SMEFT parameter space with respect to the usual differential observables and therefore can provide new constraints  to be used in global fits.

In this first quantum SMEFT tomography study, we have found that the linear interference effects of the dimension-six operators studied here ($\OO_{tG}$, $\OO_G$, $\OO_{\varphi G}$, four-fermion operators) vanish at threshold. Hence, no contribution to entanglement is present in this case. This is obviously different across the phase-space where whether the degree of entanglement is increased or decreased depends on the sign of the Wilson coefficients. This can be seen by the change of the boundaries of the entangled phase-space regions for the different contributions. 

We have also observed that the dimension-six squared contribution to the concurrence is always negative, regardless of the phase-space point. This includes the threshold and high-energy limit, where it can also be zero. Therefore, this contribution always goes in the direction of decreasing the entanglement between the top quark pair, which is most relevant when the SM produces a maximally entangled state.

We are therefore able to answer the questions posed in the introduction as follows. The line of maximal entanglement in the SM at threshold is unchanged by linear interference effects, while squared effects decrease the entanglement by inducing a triplet state. In this case, no new regions of maximal entanglement are induced by the SMEFT. However, for large values of the Wilson coefficients, the linear contributions expand the point at high energy to a wider region of phase space.

While these statements hold true when investigating the individual $gg$ and $q\bar{q}$ production channels, the picture is more involved when considering full $pp$ collisions. In this case, new physics can modify the entanglement even at the linear level, predominantly by altering the balance between the two channels.

Being exploratory, our study opens a number of questions worth being investigated. The most natural one is how much these new observables will help in better constraining top-quark SMEFT operators in the global fits, also in comparison with usual spin correlation measurements \cite{ATLAS:2019zrq,CMS:2019nrx}. Another one is whether a quantum SMEFT analysis could be used for other processes relevant for constraining new physics at present and future colliders. A promising one is the production of a pair of massive bosons which, being a bipartite qutrit system~\cite{Barr:2021zcp}, offers a richer quantum structure. 
Finally, it will be interesting to explore whether the non-trivial quantum behaviour that we have identified in the SMEFT, is the result of the employed approximations (such as tree-level computations and considering contributions up to dimension-six) or is deeply rooted and maintained at higher orders both in the gauge couplings and in the EFT expansion. 


\section*{Acknowledgements}
We would like to thank Y.~Afik and J.R.M.~de~Nova for helpful correspondence, and Eleni Vryonidou for comments on the manuscript. We would also like to thank Claudio Severi for valuable conversations and feedback.
RA's research is funded by the F.R.S-FNRS project no.\ 40005600.
EM acknowledges the support by the Minerva Foundation.
FM acknowledges the support of the F.R.S.-FNRS under the Excellence of Science EOS be.h project n.~30820817, of the IISN  and of the INFN through QFT@Colliders. 
The work of LM is supported by the European Research Council under the European Union’s Horizon 2020 research and innovation Programme (grant agreement n.950246).


\newpage
\onecolumngrid
\appendix

\newcommand*{\cVVuS}{\ensuremath{c_{VV}^{(1),u}}}
\newcommand*{\cVAuS}{\ensuremath{c_{VA}^{(1),u}}}
\newcommand*{\cAVuS}{\ensuremath{c_{AV}^{(1),u}}}
\newcommand*{\cAAuS}{\ensuremath{c_{AA}^{(1),u}}}
\newcommand*{\cVVuO}{\ensuremath{c_{VV}^{(8),u}}}
\newcommand*{\cVAuO}{\ensuremath{c_{VA}^{(8),u}}}
\newcommand*{\cAVuO}{\ensuremath{c_{AV}^{(8),u}}}
\newcommand*{\cAAuO}{\ensuremath{c_{AA}^{(8),u}}}

\section{Spin-density matrix coefficients}
\label{app:SpinDensity}

We calculate the spin density matrix \cref{eq:Rmatrix_Amplitudes} at LO analytically, including both linear and squared contributions from the SMEFT operators, using \software{FeynCalc}~9.3.1~\cite{Mertig:1990an,Shtabovenko:2016sxi,Shtabovenko:2020gxv} and generating the contributing diagrams with \software{Feynarts}~3.11~\cite{Hahn:2000kx}. We use the SMEFT@NLO~\cite{Degrande:2020evl} \software{FeynRules} model~\cite{Alloul:2013bka}.
The relevant operators for our analysis are~\cite{Degrande:2020evl}
\begin{subequations}
\label{eq:dim6_0or2F}
\begin{align}
\OO_G = g_s f^{ABC} G^{A,\mu}_\nu G^{B,\nu}_\rho G^{C,\rho}_\mu\,,\quad
\OO_{\varphi G} = \left( \varphi^\dagger \varphi - \frac{v^2}{2}\right) G_A^{\mu\nu}G^A_{\mu\nu}\,,\quad
\OO_{tG} = g_s(\bar{Q}\sigma^{\mu\nu}T^A\,t)\tilde{\varphi}G^A_{\mu\nu} + \text{h.c.}\,,
\end{align}
\end{subequations}
where $\varphi$ is the Higgs doublet, as well as the color-octet and -singlet four-fermion operators
\begin{subequations}
\label{eq:dim6_4F}
\begin{align}
\OO_{Qq}^{(8,1)} &= (\bar{Q}_L \gamma_\mu T^a Q_L)(\bar{q}_L \gamma^\mu T^a q_L)\,,\qquad
\OO_{Qq}^{(8,3)} = (\bar{Q}_L \gamma_\mu T^a \sigma^A Q_L)(\bar{q}_L \gamma^\mu T^a \sigma^A q_L)\,,\\
\OO_{tu}^{(8)} &= (\bar{t}_R \gamma_\mu T^a t_R)(\bar{u}_R \gamma^\mu T^a u_R) \,,\qquad
\OO_{td}^{(8)} = (\bar{t}_R \gamma_\mu T^a t_R)(\bar{d}_R \gamma^\mu T^a d_R)   \,,\\
\OO_{Qu}^{(8)} &= (\bar{Q}_L \gamma_\mu T^a Q_L)(\bar{u}_R \gamma^\mu T^a u_R)\,, \qquad
\OO_{Qd}^{(8)} = (\bar{Q}_L \gamma_\mu T^a Q_L)(\bar{d}_R \gamma^\mu T^a d_R)   \,,\\
\OO_{tq}^{(8)} &= (\bar{t}_R \gamma_\mu T^a t_R)(\bar{q}_L \gamma^\mu T^a q_L) \,,
\end{align}
\end{subequations}
with the corresponding singlet operators given by the same expressions but without the $SU(3)$ generators $T^a$. 
Here, $Q_L$ and $q_L$ denote heavy and light left-handed quark doublets, respectively, and $u_R$ and $d_R$ are the right-handed light quarks. 

The Fano coefficients of the expansion in \cref{eq:FanoDecomposition} are obtained from
\begin{align}
\tilde{A} = \frac{1}{4} \text{tr} (\mathds{1}_2\otimes \mathds{1}_2\,R)\,,\qquad
\tilde{B}_i^+ =  \frac{1}{4}\text{tr} (\sigma^i\otimes \mathds{1}_2 R)\,,\qquad
\tilde{B}_i^- =  \frac{1}{4}\text{tr} (\mathds{1}_2\otimes\sigma^i R)\,, \qquad
\tilde{C}_{i,j}=\frac{1}{4}\text{tr} (\sigma^i\otimes \sigma^j\,R) \,.
\end{align}
We then expand the coefficients $X=\tilde{A}$, $\tilde{C}_{ij}$ and $\tilde{B}_i^\pm$ in the new physics scale $\Lambda$ as
\begin{align}
    X = X^{(0)} + \frac{1}{\Lambda^2} X^{(1)} + \frac{1}{\Lambda^4} X^{(2)} \,,
\end{align}
where $X^{(0)}$ corresponds to the pure SM, $X^{(1)}$ is the interference between the dimension-six operators and the SM, and $X^{(2)}$ is the dimension-six squared contribution.

In order to obtain the density matrix for the full $pp$ collision, we use the luminosity functions~\cite{Bernreuther:1997gs}
\begin{align}
	L^{gg} = \frac{2 \tau}{\sqrt{s}} \int\limits_{\tau}^{1/\tau} \!\frac{d \xi}{\xi}\, f_g\left(\tau \xi\right)\, f_g\left(\frac{\tau}{\xi}\right) 
	\,,\qquad
	L^{q\bar{q}} = 2 \frac{2 \tau}{\sqrt{s}} \int\limits_{\tau}^{1/\tau} \!\frac{d \xi}{\xi}\, f_{q}\left(\tau \xi\right)\, f_{\bar{q}}\left(\frac{\tau}{\xi}\right) \,,
\end{align}
where $\xi^2 = \hat{s}/s$, $q=u,d,s,c,b$, and the $f_a(x)$ are the parton distribution functions~(PDFs).
The additional factor of two in the quark luminosity function enters as the (anti-)quark can come from either proton.
We use the \software{NNPDF4.0} NNLO PDF set~\cite{Ball:2021leu} with $\alpha_s(m_Z)=0.118$ provided by the \software{LHAPDF6}~\cite{Buckley:2014ana} interface, and evaluate the PDFs at the factorization scale $Q^2=\hat{s}$.
We assume a top quark mass of $m_t=\SI{172.76}{\GeV}$~\cite{ParticleDataGroup:2020ssz}.

\subsection{SM coefficients}
\label{app:SpinDensitySM}

For completeness, we here first present the Fano coefficients for the SM~\cite{Bernreuther:1993hq}.
In the $gg$ channel, we obtain
\begin{subequations}
\begin{align}
    \tilde{A}^{gg,(0)} &= F_{gg} \big(1 + 2\beta^2(1-z^2)-\beta^4(z^4-2z^2+2)\big),\\
    \tilde{C}^{gg,(0)}_{nn} &= - F_{gg} \big(1-2\beta^2+\beta^4(z^4-2z^2+2)\big),\\
    \tilde{C}^{gg,(0)}_{kk} &= - F_{gg} \big(1 - 2 z^2 (1-z^2) \beta^2 - (2 - 2 z^2 + z^4 ) \beta^4 \big),\\
    \tilde{C}^{gg,(0)}_{rr} &=  - F_{gg} \big(1 - (2 - 2 z^2 + z^4 )\beta^2 (2-\beta^2)  \big),\\
    \tilde{C}^{gg,(0)}_{rk} &=  F_{gg} \, 2 z\, (1-z^2)^{3/2} \beta^2 \sqrt{1-\beta^2},
\end{align}
\end{subequations}
and in the $q\bar{q}$-channel
\begin{equation}
\begin{aligned}
    \tilde{A}^{q\bar{q},(0)} = F_{q\bar{q}} \big(2+\beta^2(z^2-1)\big),\qquad
    \tilde{C}^{q\bar{q},(0)}_{nn} &= F_{q\bar{q}} \beta^2\big(z^2-1\big),\qquad
    \tilde{C}^{q\bar{q},(0)}_{kk} = F_{q\bar{q}} \big(\beta^2 + z^2 (2 - \beta^2)\big),\\
    \tilde{C}^{q\bar{q},(0)}_{rr} = F_{q\bar{q}} \big(2-\beta^2 - z^2 (2 - \beta^2)\big),&\qquad
    \tilde{C}^{q\bar{q},(0)}_{rk} = 2 z\,F_{q\bar{q}} \sqrt{(1-z^2)(1-\beta^2)},
\end{aligned}
\end{equation}
where $F_{gg} =  \frac{7+9\beta^2z^2}{192(1-\beta^2z^2)^2}$ and $F_{q\bar{q}} = \frac{1}{18}$.

\subsection{Leading-order coefficients}
\label{app:SpinDensitySMEFT_Lambda2}

We here list the linear contributions to the  Fano coefficients induced by the SMEFT operators in terms of $\beta$ and $z=\cos\theta$. The $gg$-initiated ones are given by
\begin{subequations}
\begin{align}
    \tilde{A}^{gg,(1)} &= \frac{g_s^2}{\Lambda^2} \frac{1}{1-\beta^2 z^2} \bigg[
    \frac{g_s^2 v m_t (9\beta^2z^2+7)}{12\sqrt{2}}c_{tG}
    - \frac{\beta^2m_t^4}{4m_t^2-(1-\beta^2)m_h^2}c_{\varphi G}
    + \frac{9g_s^2\beta^2m_t^2z^2}{8}c_G
    \bigg],\\
   \tilde{C}^{gg,(1)}_{nn} &= \frac{g_s^2}{\Lambda^2} \frac{1}{1-\beta^2 z^2} \bigg[
    \frac{ -7g_s^2v m_t }{12\sqrt{2}}c_{tG}
    - \frac{\beta^2m_t^4}{4m_t^2-(1-\beta^2)m_h^2}c_{\varphi G}
    + \frac{9g_s^2\beta^2m_t^2z^2}{8}c_G
    \bigg],\\
    \tilde{C}^{gg,(1)}_{kk} &= \frac{g_s^2}{\Lambda^2} \frac{1}{1-\beta^2 z^2} \bigg[
        \frac{g_s^2 v m_t \left(9 \beta ^2 z^2+7\right) \left(\beta ^2 \left(z^4-z^2-1\right)+1\right)}{12 \sqrt{2} \left(\beta ^2 z^2-1\right)} c_{tG} 
        \\\notag&\hspace{4cm}
        +\frac{\beta ^2 m_t^4}{4 m_t^2 - \left(1-\beta^2\right) m_h^2} c_{\varphi G} 
        -\frac{9 g_s^2  \beta^2 m_t^2 z^2}{8} c_G
    \bigg],\\
    \tilde{C}^{gg,(1)}_{rr} &= \frac{g_s^2}{\Lambda^2} \frac{1}{1-\beta^2 z^2} \bigg[
        \frac{g_s^2 v m_t \left(-9 \beta ^4 \left(z-z^3\right)^2-7 \beta ^2 \left(z^4-z^2+1\right)+7\right)}{12 \sqrt{2} \left(\beta ^2 z^2-1\right)} c_{tG}
        \\\notag&\hspace{4cm}
        -\frac{\beta ^2 m_t^4}{4 m_t^2-\left(1-\beta^2\right) m_h^2} c_{\varphi G}
        +\frac{9 g_s^2 \beta^2 m_t^2 z^2}{8} c_G
    \bigg], \\
    \tilde{C}^{gg,(1)}_{rk} &= \frac{g_s^2}{\Lambda^2} \frac{1}{1-\beta^2 z^2} \bigg[
        \frac{g_s^2 v m_t \beta^2 z \left(1-z^2\right) \left(9 \beta ^2+\left(\beta ^2-2\right) z^2 \left(9 \beta^2 \left(z^2-1\right)+7\right)-2\right)}{24 \sqrt{2} \sqrt{\left(\beta ^2-1\right) \left(z^2-1\right)} \left(\beta ^2 z^2-1\right)} c_{tG}
        \\\notag&\hspace{4cm}
        + \frac{9 g_s^2 \beta^2 m_t^2 z}{8} \sqrt{\frac{1-z^2}{1-\beta^2}}  c_G
    \bigg].
\end{align}
\end{subequations}
Note that at threshold ($\beta=0$) only $c_{tG}$ contributes. 
For $c_{\varphi G}$ and $c_{G}$, we have that $A^{gg,(1)} = C^{gg,(1)}_{nn}$.

For the $q\bar{q}$ channel, we organize the 7 four-fermion octet operators defined in \cref{eq:dim6_4F} in terms of their behaviour under top charge conjugation~\cite{Brivio:2019ius,Degrande:2010kt,Bernreuther:2015yna}.
The combinations of Wilson coefficients contributing in the $u\bar{u}$ or $c\bar{c}$ initiated channels are given by
\begin{equation}
\begin{aligned}
    \cVVuO &= ( c_{Qq}^{(8,1)} +  c_{Qq}^{(8,3)} + c_{tu}^{(8)} + c_{tq}^{(8)} + c_{Qu}^{(8)})/4,
    \qquad
    &\cAAuO &= ( c_{Qq}^{(8,1)} +  c_{Qq}^{(8,3)} + c_{tu}^{(8)} - c_{tq}^{(8)} - c_{Qu}^{(8)})/4, \\
    \cAVuO &= (-c_{Qq}^{(8,1)} -c_{Qq}^{(8,3)} + c_{tu}^{(8)} + c_{tq}^{(8)} - c_{Qu}^{(8)})/4,
    \quad
    &\cVAuO &= (-c_{Qq}^{(8,1)} -c_{Qq}^{(8,3)} + c_{tu}^{(8)} - c_{tq}^{(8)} + c_{Qu}^{(8)})/4,
\end{aligned}
\end{equation}
where the (axial) vector current is odd (even) under charge conjugation. 
The corresponding combinations for the $d\bar{d}$ and $s\bar{s}$ initial state are obtained by replacing  $u\to d$ in the Wilson coefficients and flipping the sign of $c_{Qq}^{(8,3)}$, whereas the color-singlet coefficients are obtained by simply replacing $(8)\to (1)$. 
In terms of these combinations, we obtain the following simple expressions for the spin-matrix coefficients for the $u\bar{u}$ and $c\bar{c}$-initiated channel
\begin{subequations}
\begin{align}
	\tilde{A}^{q\bar{q},(1)} &= \frac{4 g_s^2 m_t^2}{9 \Lambda^2 (1-\beta^2)} \bigg[
	    \sqrt{2} g_s^2 \frac{v}{m_t} (1-\beta^2) c_{tG}
	    + \left(2-(1-z^2)\beta^2\right) \cVVuO
	    + 2 z \beta \cAAuO
	\bigg],
	\\
	\tilde{C}^{q\bar{q},(1)}_{nn}
	&= - \frac{ g_s^2 m_t^2}{\Lambda^2} \frac{4 \beta^2(1-z^2)}{9(1-\beta^2)} \cVVuO ,\\
	\tilde{C}^{q\bar{q},(1)}_{kk} &= \frac{2 g_s^2 m_t^2}{9 \Lambda^2 (1-\beta^2)} \bigg[
	    2 \sqrt{2} g_s^2 \frac{v}{m_t} (1-\beta^2) z^2 c_{tG}
	    + \big(2 + \beta^2 - (2-\beta^2) (1-2 z^2) \big)\cVVuO
	    + 4 \beta z \cAAuO
	\bigg], \\
	\tilde{C}^{q\bar{q},(1)}_{rr} &= \frac{4 g_s^2 m_t^2 (1-z^2)}{9 \Lambda^2 (1-\beta^2)} \bigg[
	    \sqrt{2} g_s^2 \frac{v}{m_t} (1-\beta^2) c_{tG}
	    + (2-\beta^2) \cVVuO
	\bigg], \\
	\tilde{C}^{q\bar{q},(1)}_{rk} &= -\frac{2 g_s^2 m_t^2}{9 \Lambda^2} \sqrt{\frac{1-z^2}{1-\beta^2}} \bigg[
	    \sqrt{2} g_s^2 \frac{v}{m_t} (2-\beta^2) z c_{tG}
	    + 4 z \cVVuO
	    + 2 \beta \cAAuO
	\bigg],\\
    \tilde{B}_{k}^{\pm,q\bar{q},(1)} &= 4g_s^2\frac{m_t^2}{9\Lambda^2}\frac{1}{1-\beta^2}
    \left(
    \beta(z^2+1)\cAVuO + 2z \cVAuO
    \right),\\
    \tilde{B}_{r}^{\pm,q\bar{q},(1)} &= -4g_s^2\frac{m_t^2}{9\Lambda^2} \sqrt{\frac{1-z^2}{1-\beta^2}}
    \left(\beta z \cAVuO + 2 \cVAuO \right).
\end{align}
\end{subequations}
The corresponding coefficients for the $d\bar{d}$ and $s\bar{s}$ channel are again simply obtained by replacing $u\to d$.

We note that, at linear, order, the Fano coefficients do not receive any contributions from color-singlet operators. 
As one might expect, the Fano coefficient $\tilde{A}^{q\bar{q},(1)}$, which is proportional to the cross-section, depends only on $c_{tG}$, \cVVuO\ and \cAAuO\ at linear order.
The same is true for the $\tilde{C}^{q\bar{q},(1)}_{kk}$, $\tilde{C}^{q\bar{q},(1)}_{rr}$ and $\tilde{C}^{q\bar{q},(1)}_{rk}$, while $\tilde{C}^{q\bar{q},(1)}_{nn}$ only depends on \cVVuO.
The coefficients $\tilde{B}_i^{\pm}$, on the other hand, depend on the other two linear combinations, \cVAuO\ and \cAVuO, and only contribute to the concurrence at quadratic order, cf.\ \cref{app:Concurrence}.

It should further be noted that the expressions above assume that the quark is traveling in the positive $z$-direction. 
If instead the anti-quark is traveling in the positive $z$-direction, we need to flip $z \to -z$ (where now $z=\cos\theta$ instead of the coordinate $z$). 
In addition, we pick up an overall sign in $\tilde{C}_{rk}$ and $B_r^\pm$, as the unit vectors $\bm{n}$ and $\bm{r}$ of the helicity basis, \cref{eq:helicity_basis}, change sign~\cite{CMS:2019nrx}.
As a result, when adding the different initial-state contributions according to \cref{eq:RmatrixWeightedLuminosity}, the contributions of the operators proportional to the $c_{AA}$ and $c_{VA}$ coefficients vanish.

\subsection{Quadratic coefficients}
\label{app:SpinDensitySMEFT_Lambda4}

The dimension-six squared contributions to the $gg$-induced coefficients are
\begin{subequations}
\begin{align}
    \tilde{A}^{gg,(2)} &= \frac{m_t^4}{\Lambda^4} \frac{1}{1-\beta^2} \bigg[ 
    \frac{g_s^4 v^2 (9\beta^4z^4+4\beta^2(3z^2+4)-37)}{24 m_t^2 (\beta^2z^2-1)} c_{tG}^2
    + \frac{24 \beta^2 m_t^4}{((\beta^2-1)m_h^2+4m_t^2)^2} c_{\varphi G}^2
    \\\notag&\hspace{2.25cm}
    +\frac{27 g_s^4 (1-\beta^2z^2)}{4(1-\beta^2)}c_G^2 + \frac{2\sqrt{2}\beta^2 g_s^2 v m_t (z^2-1)}{(\beta^2z^2-1)((\beta^2-1)m_h^2 + 4m_t^2)}c_{tG}c_{\varphi G} + \frac{9 g_s^4 v}{2 m_t \sqrt{2}} c_{tG} c_{G}
    \bigg],\displaybreak[0]\\
    \tilde{C}_{nn}^{gg,(2)}
    &= \frac{m_t^4}{\Lambda^4} \frac{1}{1-\beta^2} \bigg[
        \frac{g_s^4 v^2}{m_t^2} \frac{9\beta^4z^2(z^2-2)+2\beta^2(8z^2-13)+19}{24 (\beta^2z^2-1)}c_{tG}^2
        +\frac{24\beta^2 m_t^4 }{((\beta^2-1)m_h^2 + 4 m_t^2)^2} c_{\varphi G}^2
        \\\notag&\hspace{2.25cm}
        +\frac{27 g_s^4 (\beta^2z^2-1)}{4(\beta^2-1)}c_G^2
        + \frac{2\sqrt{2}\beta^2 g_s^2 v m_t (z^2-1)}{(\beta^2z^2-1)((\beta^2-1)m_h^2+4m_t^2)} c_{tG}c_{\varphi G}
        + \frac{9 g_s^4 v}{2\sqrt{2} m_t} c_{tG}c_{G}
    \bigg],\displaybreak[0]\\
    \tilde{C}_{kk}^{gg,(2)} &= \frac{m_t^4}{\Lambda^4} \frac{1}{1-\beta^2} \bigg[
        \frac{g_s^4 v^2}{24 m_t^2} \frac{1}{(1 - \beta^2 z^2)^2} \bigg(
            9 \beta ^6 z^2 \left(z^4-2\right)
            + \beta ^4 \left(-18 z^6+25 z^4-12 z^2-14\right)
            \\\notag&\hspace{2.25cm}
            + \beta ^2 \left(-28 z^4+81 z^2+12\right)
            - 18 z^2 -19 
        \bigg) c_{tG}^2
        -\frac{24 \beta ^2 m_t^4}{\left(4 m_t^2 - m_h^2 \left(1-\beta ^2\right) \right)^2} c_{\varphi G}^2
        \\\notag&\hspace{2.25cm}
        +\frac{27 g_s^4 \left(1 - \left(2-\beta ^2\right) z^2\right)}{4 \left(1-\beta^2\right)} c_G^2
        -\frac{2 \sqrt{2} g_s^2 v m_t \beta^2 \left(1-z^2\right)}{\left(1-\beta ^2 z^2\right) \left(4 m_t^2 - m_h^2 \left(1-\beta ^2\right)\right)} c_{tG} c_{\varphi G}
        \\\notag&\hspace{2.25cm}
        + \frac{9 g_s^4 v \left(1-\left(2-\beta ^2\right) z^2\right)}{2 \sqrt{2} m_t \left(1-\beta ^2 z^2\right)} c_{tG} c_G
    \bigg], \displaybreak[0]\\
    \tilde{C}_{rr}^{gg,(2)} &= \frac{m_t^4}{\Lambda^4} \frac{1}{1-\beta^2} \bigg[
        \frac{g_s^4 v^2}{24 m_t^2} \frac{1}{(1 - \beta^2 z^2)^2} \bigg(
            -9 \beta ^6 z^2 \left(z^4-2 z^2+2\right)
            + \beta ^4 \left(18 z^6-57 z^4+52 z^2-14\right)
            \\\notag&\hspace{2.25cm}
            + \beta ^2 \left(28 z^4-57 z^2+58\right)
            +18 z^2-37
        \bigg) c_{tG}^2
        + \frac{24 \beta^2 m_t^4}{\left(4 m_t^2 - m_h^2 \left(1-\beta ^2\right)\right)^2} c_{\varphi G}^2
        \\\notag&\hspace{2.25cm}
        - \frac{27 g_s^4 \left(1-\left(2-\beta ^2\right) z^2\right)}{4 \left(1-\beta ^2\right)} c_G^2
        + \frac{2 \sqrt{2} g_s^2 v m_t \beta^2 \left(1-z^2\right)}{\left(1-\beta ^2 z^2\right) \left(4 m_t^2 - m_h^2 \left(1-\beta ^2\right)\right)} c_{tG} c_{\varphi G}
        \\\notag&\hspace{2.25cm}
        - \frac{9 g_s^4 v \left(1-\left(2-\beta ^2\right) z^2\right)}{2 \sqrt{2} m_t \left(1-\beta ^2 z^2\right)} c_{tG} c_{G}
    \bigg],\displaybreak[0] \\
    \tilde{C}_{rk}^{gg,(2)} &= \frac{m_t^4}{\Lambda^4} \frac{1}{\sqrt{1-\beta^2}} \bigg[
        \frac{g_s^4 v^2}{192 m_t^2} \frac{1}{(1 - \beta^2 z^2)^2} \bigg(
            - 144 \beta^4 z^3 \left(1-z^2\right)^{3/2}
            + 16 \beta^2 z \sqrt{1-z^2} \left(14 z^2-23\right) 
            \\\notag&\hspace{2.25cm}
            + 144 z \sqrt{1-z^2} 
        \bigg) c_{tG}^2
        + \frac{27}{2} g_s^4 z\, \frac{\sqrt{1-z^2}}{1-\beta^2} c_{G}^2
        - \frac{2 \sqrt{2} g_s^2 v m_t \beta^2 z \sqrt{1-z^2}}{\left(1-\beta ^2 z^2\right) \left(4 m_t^2 - m_h^2 \left(1-\beta ^2\right)\right)} c_{tG} c_{\varphi G}
        \\\notag&\hspace{2.25cm}
        +\frac{9 g_s^4 v z \sqrt{1-z^2}}{\sqrt{2} m_t \left(1-\beta ^2 z^2\right)} c_{tG} c_G
    \bigg],
\end{align}
\end{subequations}
while for $q\bar{q}$ we have
\begin{subequations}
\begin{align}
    \tilde{A}^{q\bar{q},(2)} &= \frac{4 m_t^4}{9(1-\beta^2)^2\Lambda^4} \bigg[
        \frac{g_s^4 v^2}{m_t^2} \left(1-\beta ^2\right)  \left(2-\beta ^2 \left(z^2+1\right)\right) c_{tG}^2
        +\frac{g_s^2 v}{m_t} 4 \sqrt{2} \left(1-\beta ^2\right) (\cVVuO + \beta z \cAAuO ) c_{tG}
        \\\notag&\phantom{=}\ 
        + \beta ^2 \bigg(9 (\cAAuS)^2 \left(z^2+1\right)
        +2 (\cAAuO)^2 \left(z^2+1\right)
        +z^2 \left(9 (\cAVuS)^2+2 (\cAVuO)^2+9 (\cVAuS)^2+2 (\cVAuO)^2
        \right.\\\notag&\phantom{=}\ \left.+9 (\cVVuS)^2+2 (\cVVuO)^2\right)
        +9 (\cAVuS)^2+2 (\cAVuO)^2-9 (\cVAuS)^2-2 (\cVAuO)^2 -9 (\cVVuS)^2 -2 (\cVVuO)^2\bigg)
        \\\notag&\phantom{=}\
        +4 \beta  z (9 \cAAuS \cVVuS+2 \cAAuO \cVVuO+9 \cAVuS \cVAuS+2 \cAVuO \cVAuO)
        \\\notag&\phantom{=}\
        +18 (\cVAuS)^2+4 (\cVAuO)^2+18 (\cVVuS)^2+4 (\cVVuO)^2
    \bigg] ,
    \displaybreak[0]\\
    \tilde{C}_{nn}^{q\bar{q},(2)} &=  \frac{4 m_t^4}{9 \Lambda^4} \frac{\beta^2 ( 1-z^2)}{(1-\beta^2)^2}\bigg[
        \frac{g_s^4 v^2}{m_t^2} \left(1-\beta ^2\right) c_{tG}^2
        + 9 (\cAAuS)^2+2 (\cAAuO)^2+9 (\cAVuS)^2+2 (\cAVuO)^2-9 (\cVAuS)^2
        \\\notag&\phantom{=}\
        -2 (\cVAuO)^2-9 (\cVVuS)^2-2 (\cVVuO)^2
    \bigg] ,
    \displaybreak[0]\\
    \tilde{C}_{kk}^{q\bar{q},(2)} &=  \frac{4 m_t^4}{9 (1-\beta^2)^2\Lambda^4} \bigg[
        \frac{g_s^4 v^2}{m_t^2} \left(1-\beta ^2\right) \left( z^2(2-\beta^2)-\beta^2 \right) c_{tG}^2
        + \frac{g_s^2 v}{m_t} 4 \sqrt{2} \left(1-\beta ^2\right) z \left(\beta \cAAuO + z \cVVuO\right) c_{tG}
        \\\notag&\phantom{=}\
        + \beta ^2 \left(
            9 (\cAAuS)^2 \left(z^2+1\right)+2 (\cAAuO)^2 \left(z^2+1\right)
            +z^2 \left(9 (\cAVuS)^2+2 (\cAVuO)^2-9 (\cVAuS)^2-2 (\cVAuO)^2
            \right.\right.\\\notag&\phantom{=}\ \left.\left.
            -9 (\cVVuS)^2-2 (\cVVuO)^2\right)
            +9 (\cAVuS)^2+2 (\cAVuO)^2+9 (\cVAuS)^2+2 (\cVAuO)^2+9 (\cVVuS)^2+2 (\cVVuO)^2
        \right)
        \\\notag&\phantom{=}\
        +4 \beta  z (9 \cAAuS \cVVuS+2 \cAAuO \cVVuO+9 \cAVuS \cVAuS+2 \cAVuO \cVAuO)
        \\\notag&\phantom{=}\
        +2 z^2 \left(9 (\cVAuS)^2+2 (\cVAuO)^2+9 (\cVVuS)^2+2 (\cVVuO)^2\right)
    \bigg] ,
    \displaybreak[0]\\
    \tilde{C}_{rr}^{q\bar{q},(2)} &=  \frac{4 m_t^4}{9 \Lambda^4} \frac{1-z^2}{(1-\beta^2)^2} \bigg[
        \frac{g_s^4 v^2}{m_t^2} ( 2 - 3 \beta^2 + \beta^4 ) c_{tG}^2
        + \frac{g_s^2 v}{m_t} 4 \sqrt{2} \left(1-\beta^2\right) \cVVuO c_{tG}
        \\\notag&\phantom{=}\
        -\beta ^2 \left( 
            9 (\cAAuS)^2+2 (\cAAuO)^2
            +9 (\cAVuS)^2+2 (\cAVuO)^2+2 (\cVAuO)^2+9 (\cVVuS)^2+2 (\cVVuO)^2
        \right)
        \\\notag&\phantom{=}\
        -9 \left(\beta ^2-2\right) (\cVAuS)^2
        +4 (\cVAuO)^2+18 (\cVVuS)^2+4 (\cVVuO)^2
    \bigg] ,
    \displaybreak[0]\\
        \tilde{C}_{rk}^{q\bar{q},(2)} &= -\frac{8 m_t^4}{9 (1-\beta^2) \Lambda^4} \sqrt{\frac{1-z^2}{1-\beta^2}}\,\bigg[
            \frac{g_s^4 v^2}{m_t^2} z (1-\beta^2) c_{tG}^2
            + \frac{g_s^2 v}{m_t} \sqrt{2} \left( z (2-\beta^2) \cVVuO + \beta \cAAuO \right) c_{tG}
            \\\notag&\phantom{=}\
            + \beta  \left(9 \cAAuS \cVVuS+2 \cAAuO \cVVuO+9 \cAVuS \cVAuS+2 \cAVuO \cVAuO\right)
            \\\notag&\phantom{=}\
            +z \left(9 (\cVAuS)^2+2 (\cVAuO)^2+9 (\cVVuS)^2+2 (\cVVuO)^2\right)
        \bigg] ,
\end{align}
\end{subequations}
where, again, the contributions above are for up-type quarks, whereas the down-type contribution is obtained replacing $u\to d$.
At the quadratic order,  all the combinations \cVVuO, \cAAuO, \cVAuO\ and \cAVuO\ as well as their singlet counterparts contribute.

\section{Angular-averaged coefficients}
\label{app:SpinDensityOmega}

We now execute the integration over the solid angle according to \cref{eq:AverageRmatrix}, switching from the (top-direction dependent) helicity basis \cref{eq:helicity_basis} to the beam basis.
Averaging over the azimuthal angle, the correlation matrix $C_{ij}$ becomes diagonal with two identical entries, $\tilde{C}_\perp \equiv \tilde{C}_{xx}=\tilde{C}_{yy}$, and the only non-vanishing component of the polarization vectors is $B_z^\pm$. 
We are then left with the integrals
\begin{equation}
\begin{aligned}
    \tilde{A}_\Omega(\beta) &= \frac{1}{2} \int\limits_{-1}^{1} dz\,  \tilde{A}(\beta,z) \,,
    \qquad
    \tilde{B}_{z,\Omega}^\pm(\beta) =  \frac{1}{2} \int\limits_{-1}^{1} dz\, \left[ 
        z\, \tilde{B}_k^\pm(\beta,z) + \sqrt{1-z^2} \,\tilde{B}_r^\pm(\beta,z)
    \right],
    \\
    \tilde{C}_{z,\Omega}(\beta) &=  \frac{1}{2} \int\limits_{-1}^{1} dz\, \left[
        z^2\, \tilde{C}_{kk}(\beta,z) + (1-z^2)\, \tilde{C}_{rr}(\beta,z) + 2\, z \sqrt{1-z^2} \,\tilde{C}_{rk}(\beta,z)
    \right],\\
    \tilde{C}_{\perp,\Omega}(\beta) &=  \frac{1}{2} \int\limits_{-1}^{1} dz\, \left[
        \tilde{C}_{nn}(\beta,z)
        + (1-z^2) \,\tilde{C}_{kk}(\beta,z)
        + z^2\, \tilde{C}_{rr}(\beta,z)
       - 2 \,z \sqrt{1-z^2}\, \tilde{C}_{rk}(\beta,z)
    \right].
\end{aligned}
\end{equation}

In the SM, we have~\cite{Afik:2020onf}
\begin{subequations}
\begin{align}
    \tilde{A}^{gg,(0)}_\Omega &= \frac{1}{192} \left[  
        -59 + 31 \beta^2
        + 2 ( 33 - 18 \beta^2 + \beta^4 ) \frac{\artanh\beta}{\beta}
    \right] , \displaybreak[0]\\
    \tilde{C}^{gg,(0)}_{z,\Omega} &= \frac{1}{2880 \beta^4}\bigg[
        879 \beta^6 - 3413 \beta^4 + 4450 \beta^2 - 2940 
        + 4 ( 72 \beta^4 - 745 \beta^2 + 735 ) \sqrt{1-\beta^2}
        \\\notag&\hspace{2cm}
        + 30 \left( \beta^8 -53 \beta^6 + 151 \beta^4 - 181 \beta^2 + 98  - 2 ( 17 \beta^4 -66\beta^2 + 49 ) \sqrt{1-\beta^2} \right) \frac{\artanh\beta}{\beta}
    \bigg],\displaybreak[0]\\
    \tilde{C}^{gg,(0)}_{\perp,\Omega} &=  \frac{1}{2880 \beta^4}\bigg[
        -207 \beta^6 + 1024 \beta^4 - 2225 \beta^2 + 1470
        - 2 ( 72 \beta^4 - 745 \beta^2 + 735 ) \sqrt{1-\beta^2}
        \\\notag&\hspace{2cm}
        + 15 \left( 33 \beta^6 - 116 \beta^4 + 181 \beta^2 - 98  + 2 ( 17 \beta^4 -66\beta^2 + 49 ) \sqrt{1-\beta^2} \right) \frac{\artanh\beta}{\beta}
    \bigg],
\end{align}
\end{subequations}
in the $gg$ channel, and in the $q\bar{q}$ channel we obtain
\begin{align}
    \tilde{A}^{q\bar{q},(0)}_\Omega &=  \frac{3-\beta^2}{27}\,,  &\qquad
    \tilde{C}^{q\bar{q},(0)}_{z,\Omega} &= \frac{11 - 3 \beta^2 + 4 \sqrt{1-\beta^2}}{135} \,,&\qquad
    \tilde{C}^{q\bar{q},(0)}_{\perp,\Omega} &=  \frac{2 - \beta^2 - 2 \sqrt{1-\beta^2}}{135}\,.
\end{align}

The SMEFT contributions at linear order are given by
\begin{subequations}
\begin{align}
    \tilde{A}^{gg,(1)}_\Omega &=  -\frac{g_s^2 m_t^2}{\Lambda^2} \left[
        \frac{g_s^2 v}{m_t} \frac{9\beta - 16 \artanh\beta}{12 \sqrt{2} \beta} c_{tG}
        + \frac{m_t^2 \beta \artanh\beta}{4 m_t^2 - m_h^2 (1-\beta^2)} c_{\varphi G}
        + 9 g_s^2 \frac{\beta-\artanh\beta}{8 \beta} c_{G}
    \right],  \displaybreak[0]\\
    \tilde{C}^{gg,(1)}_{z,\Omega} &= - \frac{m_t^2}{\Lambda^2} \bigg[ 
        \frac{\sqrt{2}  g_s^4 v}{360\, m_t} \bigg( 
            \frac{1470 -1865 \beta^2 + 529 \beta^4 - 72 \beta^6}{\beta^4 \sqrt{1-\beta^2}}
            + \frac{1470 - 1130 \beta^2 + 264 \beta^4}{\beta^4}
            \\\notag&\hspace{1.5cm}
            + 15 \frac{9 \beta^6 - 68 \beta^4 + 157 \beta^2 -98 + \left(\beta^6 -27 \beta^4 + 108 \beta^2 -98\right)\sqrt{1-\beta^2}}{\beta^4 \sqrt{1-\beta^2}}  \frac{\artanh\beta}{\beta}
        \bigg) c_{tG} 
        \\\notag&\hspace{1.5cm}
        + \frac{g_s^2 m_t^2}{\beta} \frac{2 \beta - (2-\beta^2) \artanh\beta}{4 m_t^2 - m_h^2 (1-\beta^2)} c_{\varphi G}
        \\\notag&\hspace{1.5cm}
        + \frac{3 g_s^4}{8 \beta^3} \frac{4 \beta^3 - 6 \beta + (\beta^3 - 6 \beta)\sqrt{1-\beta^2} + \left(6 - 6 \beta^2 + 3 (2-\beta^2)\sqrt{1-\beta^2}\right)\artanh\beta}{\sqrt{1-\beta^2}} c_G
    \bigg],\displaybreak[0]\\
    \tilde{C}^{gg,(1)}_{\perp,\Omega} &= \frac{m_t^2}{\Lambda^2} \bigg[
        \frac{\sqrt{2} g_s^4 v}{720 m_t \beta^4 (1-\beta^2)} \bigg(
            111 \beta^6 + 1019 \beta^4 -2600 \beta^2 + 1470 
            \\\notag&\hspace{1.5cm}
            - (72 \beta^6 - 529 \beta^4 + 1865 \beta^2 - 1470) \sqrt{1-\beta^2} 
            - 15 \left(1-\beta^2\right)\left( \beta^6 + 23 \beta^4 -108 \beta^2 + 98 \right) \frac{\artanh\beta}{\beta}
            \\\notag&\hspace{1.5cm}
            - 15 \left( 9 \beta^4 - 59 \beta^2 + 98 \right) \left(1-\beta^2\right)^{3/2} \frac{\artanh\beta}{\beta}
        \bigg) c_{tG}
        + \frac{g_s^2 m_t^2 \left(\beta - \artanh\beta\right)}{\beta \left(4 m_t^2 - m_h^2 (1-\beta^2)\right)} c_{\varphi G}
        \\\notag&\hspace{1.5cm}
        + \frac{3 g_s^4}{8} \frac{2 \beta^3 - 3 \beta - (\beta^3 +3\beta)\sqrt{1-\beta^2} + 3 \left( 1-\beta^2 +\sqrt{1-\beta^2} \right)\artanh\beta}{\beta^3 \sqrt{1-\beta^2}} c_G 
    \bigg],
\end{align}
\end{subequations}
\begin{subequations}
\begin{align}
    \tilde{A}^{q\bar{q},(1)}_\Omega &= \frac{4 g_s^2 m_t^2}{27 \Lambda^2} \left[ 3 \sqrt{2}\, \frac{g_s^2 v}{m_t}\, c_{tG} + 2 \frac{3-\beta^2}{1-\beta^2} \,\cVVuO\, \right], \displaybreak[0]\\
    \tilde{C}^{q\bar{q},(1)}_{z,\Omega} &= \frac{8 g_s^2 m_t^2}{135 \Lambda^2} \left[ 
        \frac{g_s^2 v}{m_t} \frac{2 \beta^2-4+11 \sqrt{1-\beta^2}}{\sqrt{2-2 \beta^2}}\, c_{tG}+\frac{11-3 \beta^2-4 \sqrt{1-\beta^2}}{1-\beta^2} \,\cVVuO
    \right], \displaybreak[0]\\
    \tilde{C}^{q\bar{q},(1)}_{\perp,\Omega} &= \frac{8 g_s^2 m_t^2}{135 \Lambda ^2} \left[
        \frac{g_s^2 v}{m_t} \frac{2-\beta ^2+2 \sqrt{1-\beta ^2}}{\sqrt{2-2 \beta ^2}} c_{tG}
        + \frac{2-\beta ^2+2 \sqrt{1-\beta ^2}}{1-\beta ^2} \cVVuO
    \right], \displaybreak[0]\\
    B_{z,\Omega}^{q\bar{q},(1)} &= \frac{8 g_s^2 m_t^2}{27 \Lambda ^2} \frac{1-2 \sqrt{1-\beta ^2}}{1-\beta ^2} \cVAuO
\end{align}
\end{subequations}
and the dimension-six squared corrections are
\begin{subequations}
\begin{align}
    \tilde{A}^{gg,(2)}_\Omega &= \frac{m_t^4}{\Lambda^4} \frac{1}{1-\beta^2} \bigg[
        \frac{g_s^4 v^2}{m_t^2} \frac{3 \beta (7+\beta^2) + 16 (1-\beta^2) \artanh\beta}{24 \beta} c_{tG}^2
        + \frac{24 m_t^4 \beta^2}{\left(4 m_t^2 - m_h^2 (1-\beta^2)\right)^2} c_{\varphi G}^2
        \\\notag&\hspace{2.5cm}
        + \frac{9 g_s^4}{4} \frac{3-\beta^2}{1-\beta^2} c_G^2
        + 2 \sqrt{2} g_s^2 m_t v \frac{\beta - (1-\beta^2)\artanh\beta}{\beta \left(4 m_t^2 - m_h^2 (1-\beta^2)\right)} c_{tG} c_{\varphi G}
        + \frac{9 g_s^4 v}{2 \sqrt{2} m_t} c_{tG} c_G  
    \bigg], \displaybreak[0]\\
    \tilde{C}^{gg,(2)}_{z,\Omega} &= \frac{m_t^4}{\Lambda^4} \frac{1}{1-\beta^2} \bigg[
        \frac{g_s^4 v^2}{360 m_t^2 \beta^4} \bigg(
            429 \beta^6 - 2773 \beta^4 + 7100 \beta^2 -4800
            - \left(288 \beta^4 - 4700 \beta^2 + 4800 \right) \sqrt{1-\beta ^2}
            \\\notag&\phantom{=}\
            + 30 \left(\beta^8 - 12 \beta^6 +141 \beta^4 -290 \beta^2 + 160 + 10 \left(5 \beta^4-21 \beta^2+16\right) \sqrt{1-\beta ^2} \right) \frac{\artanh\beta}{\beta}
        \bigg) c_{tG}^2
        \\\notag&\phantom{=}\
        + \frac{8 m_t^4 \beta ^2}{\left(4 m_t^2 - m_h^2(1-\beta^2)\right)^2} c_{\varphi G}^2
        + \frac{9 g_s^4}{20} \frac{\beta^2-7+8\sqrt{1-\beta^2}}{1-\beta^2} c_G^2
        -\frac{2 \sqrt{2} g_s^2 m_t v}{3 \beta ^3 \sqrt{1-\beta ^2}} \bigg(
            \frac{4 \beta^5 - 10 \beta^3 + 6 \beta}{4 m_t^2 - (1-\beta^2) m_h^2}
            \\\notag&\phantom{=}\
            + \frac{ 6 \beta -7\beta^3 }{4 m_t^2 - (1-\beta^2) m_h^2} \sqrt{1-\beta ^2}
            - 3 \frac{2 (1-\beta^2)^2 + (\beta^4 -3 \beta^2 + 2) \sqrt{1-\beta ^2}}{4 m_t^2 - (1-\beta^2) m_h^2} \artanh\beta
        \bigg) c_{tG} c_{\varphi G}
        \\\notag&\phantom{=}\
        + \frac{3 \sqrt{2} g_s^4 v}{4 m_t \beta^5 \sqrt{1-\beta^2}} \bigg(
            8 \beta^5 -20 \beta^3 + 12 \beta
            + ( \beta^5 - 14 \beta^3 + 12 \beta) \sqrt{1-\beta^2} 
            \\\notag&\phantom{=}\
            - 6 (1-\beta^2) \left( 2 (1-\beta^2) + (2-\beta^2) \sqrt{1-\beta^2} \right) \artanh\beta
        \bigg) c_{tG} c_G
    \bigg],  
    \displaybreak[0]\\
    \tilde{C}^{gg,(2)}_{\perp,\Omega} &= \frac{m_t^4}{\Lambda^4} \frac{1}{1-\beta^2} \bigg[
        \frac{g_s^4 v^2}{360 m_t^2 \beta^4} \bigg(
            48 \beta^6+ 1094 \beta^4 - 3550 \beta^2 + 2400
            + \left(144 \beta ^4-2350 \beta ^2+2400 \right) \sqrt{1-\beta ^2}
            \\\notag&\phantom{=}\
            + 15 \left( \beta^8 + 28 \beta^6 - 159 \beta^4 + 290 \beta^2 - 160 - 10 \left(5 \beta^4 - 21 \beta^2 + 16\right) \sqrt{1-\beta ^2}\right) \frac{\artanh\beta}{\beta}
        \bigg)c_{tG}^2
        \\\notag&\phantom{=}\
        + \frac{8 m_t^4 \beta ^2}{\left(4 m_t^2 - m_h^2(1-\beta^2)\right)^2} c_{\varphi G}^2
        - \frac{9 g_s^4}{20} \frac{3 \beta^2-11+4\sqrt{1-\beta^2}}{1-\beta^2} c_G^2
        \\\notag&\phantom{=}\
        - \frac{2\sqrt{2} g_s^2 m_t v}{3 \beta^2 \left(4 m_t^2 - m_h^2(1-\beta^2)\right)} \left(1 + \sqrt{1-\beta^2}\right) \left(
            2 \beta^2 - 3 + 3 (1-\beta^2) \frac{\artanh\beta}{\beta}
        \right) c_{tG} c_{\varphi G}
        \\\notag&\phantom{=}\
        + \frac{3 \sqrt{2} g_s^4 v}{4 m_t \beta^4} \bigg(
            \beta^4 + 7 \beta^2 - 6 
            + (4\beta^2-6)\sqrt{1-\beta^2}
            + 3 \left(
                2 - 3 \beta^2 + \beta^4
                + 2 (1-\beta^2)^{3/2}
            \right) \frac{\artanh\beta}{\beta}
        \bigg) c_{tG} c_G
    \bigg],
\end{align}
\end{subequations}
\begin{subequations}
\begin{align}
    \tilde{A}^{q\bar{q},(2)}_\Omega &= \frac{8 m_t^4}{27 \Lambda^4} \bigg[ 
        \frac{g_s^4 v^2}{m_t^2} \frac{3 - 2 \beta^2}{1-\beta^2}\, c_{tG}^2
        + 6 \sqrt{2} \, \frac{g_s^2 v}{m_t} \frac{1}{1-\beta^2}\, c_{tG} \cVVuO \\
        &\hspace{2cm}+ \frac{3-\beta^2}{(1-\beta^2)^2} \left( 9 (\cVVuS)^2 + 9 (\cVAuS)^2 + 2 (\cVVuO)^2 + 2 (\cVAuO)^2\right) \notag\\
        &\hspace{2cm}+ \frac{2\,\beta^2}{(1-\beta^2)^2} \left( 9 (\cAAuS)^2 + 9 (\cAVuS)^2 + 2 (\cAAuO)^2 + 2 (\cAVuO)^2\right)
    \bigg], \notag \displaybreak[0]\\
    \tilde{C}^{q\bar{q},(2)}_{z,\Omega} &= \frac{8 m_t^4}{135 \Lambda ^4} \frac{1}{\left(1-\beta ^2\right)^2} \bigg[
        \frac{g_s^4 v^2 }{m_t^2} (1-\beta^2) \left(11-8 \beta^2-4 \sqrt{1-\beta ^2}\right) c_{tG}^2 
        \\&\hspace{3.25cm}
        + 2 \sqrt{2} \frac{g_s^2 v}{m_t} \left(11 \left(1-\beta ^2\right)+\left(2 \beta ^2-4\right) \sqrt{1-\beta ^2}\right) c_{tG} \cVVuO
        \notag\\&\hspace{3.25cm}
        + \left(11 -3 \beta ^2-4 \sqrt{1-\beta ^2}\right) \left( 9 (\cVVuS)^2 + 9 (\cVAuS)^2 + 2 (\cVVuO)^2 + 2 (\cVAuO)^2 \right)
    \bigg], \notag \displaybreak[0]\\
    \tilde{C}^{q\bar{q},(2)}_{\perp,\Omega} &= \frac{8 m_t^4}{135 \Lambda ^4} \frac{1}{\left(1-\beta ^2\right)^2} \bigg[
        \frac{g_s^4 v^2 }{m_t^2} (1-\beta^2) \left(2-\beta ^2+2 \sqrt{1-\beta ^2}\right) c_{tG}^2
        \\&\hspace{3.25cm}
        + 2 \sqrt{2} \frac{g_s^2 v}{m_t} \left(2 \left(1-\beta ^2\right)+\left(2-\beta ^2\right) \sqrt{1-\beta ^2}\right) c_{tG} \cVVuO
        \notag\\&\hspace{3.25cm}
        + 5 \beta^2 \left( 9 (\cAAuS)^2 + 9 (\cAVuS)^2 + 2 (\cAAuO)^2 + 2 (\cAVuO)^2 \right)
        \notag\\&\hspace{3.25cm}
        + \left(2-\beta^2 + 2 \sqrt{1-\beta^2}\right) \left( 9 (\cVVuS)^2 + 9 (\cVAuS)^2 + 2 (\cVVuO)^2 + 2 (\cVAuO)^2 \right)
    \bigg]. \notag
\end{align}
\end{subequations}

\section{Concurrence}
\label{app:Concurrence}

Let us now take the SM plus dimension-six operators and, for simplicity, work in the basis where $C_{ij}$ is diagonal. 
The eigenvalues of $C_{ij}$ and the elements $B_i^\pm$ can be expanded as
\begin{subequations}\begin{align}
    C_i &= C_i^{(0)} + \frac{1}{\Lambda^2}  C_i^{(1)} + \frac{1}{\Lambda^4}  C_i^{(2)} \,\qquad \text{and} \qquad
    B_i =  \frac{1}{\Lambda^2}  B_i^{(1)} + \frac{1}{\Lambda^4}  B_i^{(2)} \,, 
\end{align}\end{subequations}
where we used that $B_i^\pm=0$ in the SM and that, in the absence of $CP$-odd operators, $B_i^+ = B_i^-$.  Defining
\begin{subequations}\begin{align}
    \bar{I} &= 1 - \frac{1}{\Lambda^4} \left[ \frac{{B_1^{(1)}}^2}{1 + C_1^{(0)}} + \frac{{B_2^{(1)}}^2}{1 + C_2^{(0)}} \right] \,,\\
    \bar{C}_i &= C_i^{(0)} + \frac{1}{\Lambda^2} C_i^{(1)} + \frac{1}{\Lambda^4} C_i^{(2)} - \frac{1}{\Lambda^4} \frac{{B_i^{(1)}}^2}{1 + C_i^{(0)}}\,,
\end{align}\end{subequations}
the eigenvalues of $\omega$ are then given by $\bar{I}-\bar{C}_3 \pm (\bar{C}_1 + \bar{C}_2)$ and $\bar{I}+\bar{C}_3 \pm (\bar{C}_1 - \bar{C}_2)$, and the concurrence simply reads
\begin{equation}
    C[\rho] = \max\left(0, \frac{\pm \bar{C}_3 + \left|\bar{C}_1\mp\bar{C}_2\right| - \bar{I}}{2}\right)\,.
\end{equation}
Further using that, in this case, the $\hat{n}$ direction is uncorrelated with the other two directions, {\it  i.e.},\ $C_{nk}=C_{kn}=C_{nr}=C_{rn}=0$, as well as that the polarization in this direction vanishes, {\it i.e.},\ $B^\pm_n=0$, we see that also $B_3=0$ in the diagonal basis. 
Including the dimension-six-squared contributions, we obtain
\begin{subequations}\begin{align}
    C_{1,2}^{(0)} &= \frac{1}{2} \left[ C_{rr}^{(0)} + C_{kk}^{(0)} \pm \sqrt{\left(C_{rr}^{(0)} - C_{kk}^{(0)}\right)^2 + 4\, {C_{rk}^{(0)}}^2} \right] \,, \displaybreak[0]\\
    C_{1,2}^{(1)} &= \frac{1}{2} \left(
        C_{rr}^{(1)} + C_{kk}^{(1)} 
        \pm \frac{\left(C_{rr}^{(1)} - C_{kk}^{(1)}\right) \left(C_{rr}^{(0)} - C_{kk}^{(0)}\right) + 4\,C_{rk}^{(1)} C_{rk}^{(0)} }{\sqrt{\left(C_{rr}^{(0)} - C_{kk}^{(0)}\right)^2 + 4\, {C_{rk}^{(0)}}^2}}
    \right)\,, \displaybreak[0]\\
    C_{1,2}^{(2)} &= \frac{1}{2} \left(
        C_{rr}^{(2)} + C_{kk}^{(2)} 
        \pm \frac{\left(C_{rr}^{(2)} - C_{kk}^{(2)}\right) \left(C_{rr}^{(0)} - C_{kk}^{(0)}\right) + 4\,C_{rk}^{(2)} C_{rk}^{(0)} }{\sqrt{\left(C_{rr}^{(0)} - C_{kk}^{(0)}\right)^2 + 4\, {C_{rk}^{(0)}}^2}}
    \right) 
    \\ \notag
    &\hspace{1cm} 
    \pm \frac{\left[
        \left(C_{rr}^{(1)} - C_{kk}^{(1)}\right) C_{rk}^{(0)} - C_{rk}^{(1)} \left(C_{rr}^{(0)} - C_{kk}^{(0)}\right)
    \right]^2}{\left[\left(C_{rr}^{(0)} - C_{kk}^{(0)}\right)^2 + 4\, {C_{rk}^{(0)}}^2\right]^{3/2}} \,,\displaybreak[0]\\
    C_{3}^{(0,1,2)} &= C_{nn}^{(0,1,2)} \,,
    \intertext{as well as}
    B_1^{(1)} &= \frac{B_r^{(1)}\,2\,C_{rk}^{(0)} + B_k^{(1)} \left(C_{rr}^{(0)}-C_{kk}^{(0)}+\sqrt{\left(C_{rr}^{(0)} - C_{kk}^{(0)}\right)^2 + 4\, {C_{rk}^{(0)}}^2}\right)}{\sqrt{4\,{C_{rk}^{(0)}}^2+\left(C_{rr}^{(0)}-C_{kk}^{(0)}+\sqrt{\left(C_{rr}^{(0)} - C_{kk}^{(0)}\right)^2 + 4\, {C_{rk}^{(0)}}^2}\right)^2}}\,, \displaybreak[0]\\
    B_2^{(1)} &= \frac{B_r^{(1)} \left(C_{rr}^{(0)}-C_{kk}^{(0)}+\sqrt{\left(C_{rr}^{(0)} - C_{kk}^{(0)}\right)^2 + 4\, {C_{rk}^{(0)}}^2}\right) - B_k^{(1)}\,2\,C_{rk}^{(0)}}{\sqrt{4\,{C_{rk}^{(0)}}^2+\left(C_{rr}^{(0)}-C_{kk}^{(0)}+\sqrt{\left(C_{rr}^{(0)} - C_{kk}^{(0)}\right)^2 + 4\, {C_{rk}^{(0)}}^2}\right)^2}}\,.
\end{align}\end{subequations}

\section{Probabilities}
\label{app:Probabilities}
In~\cref{sec:QuantumState}, we have defined the probabilities to find the top quark pair in a triplet state. 
As a function of the Wilson coefficients, these are
\begin{align}
    p_{gg} &= \frac{72}{7\Lambda^4} m_t^2 (3\sqrt{2} m_t \, c_G + v \, c_{tG})^2\,,\\
    p_{q\bar q} &= \frac{1}{2}-4\frac{\cVAuO}{\Lambda^2}+ \frac{8m_t^4}{\Lambda^4}
    \big( \frac{v\sqrt{2}}{m_t} \cVAuO c_{tG} 
    - 9 \cVAuS \cVVuS
    + 2 \cVAuO \cVVuO
    \big),
\end{align}
at threshold, while for the high-energy central region ($z=0$) we have
\begin{align}
    p_{\Psi^+}^{gg} &= \frac{\beta^2}{1+2\beta^2-2\beta^4} + 
    \frac{4\beta^2m_t}{\Lambda^2(1-2\beta^2-2\beta^4)^3}
    \bigg(7\sqrt{2}v (1-4\beta^4(1-\beta^2)^2)c_{tG} +
    \frac{48\beta^2m_t^3(1+2\beta^2-2\beta^4)}{g_s^2(1-\beta^2)m_h^2+4m_t^2}c_{\varphi G}\bigg)\\
    &+\frac{-8\beta^2m_t^2}{49(1-\beta^2)^2(1-2\beta^2-2\beta^4)^3}
    \bigg(
    \frac{576\beta^2(\beta^2-1)(6\beta^4-6\beta^2-7)m_t^6}{g_s^4((\beta^2-1)m_h^2+4m_t^2)^2}c_{\varphi G}^2\notag\\
    &+\frac{336\sqrt{2}\beta^2(\beta^2-1)(4\beta^6-6\beta^4+8\beta^2-7)v m_t^3}{g_s^2((\beta^2-1)m_h^2+4m_t^2)}c_{tG}c_{\varphi G}\notag\\
    &- 7(162(2\beta^4-2\beta^2-1)c_G^2m_t^2 -54 \sqrt{2}(2\beta^6-4\beta^4+\beta^2+1) v m_t c_{G}c_{tG}  \notag\\\nonumber
    &+(28\beta^{12} -112 \beta^{10}+284\beta^8 +506 \beta^6 + 507 \beta^4 - 231 \beta^2 + 26
    ) v^2 c_{tG}^2
    \bigg),
    \\
    p_{\Psi^+}^{uu} &= \frac{1}{2-\beta^2} - \frac{4m_tv\sqrt{2}\beta^2}{(2-\beta^2)^2\Lambda^2}c_{tG} 
    +\frac{8\beta^2m_t^2}{(2-\beta^2)^3(1-\beta^2)^2g_s^4\Lambda^4}
    \bigg(
        (\beta^6+3\beta^2-8\beta^2+4) g_s^4 v^2 c_{tG}^2
        \\\notag&\phantom{=}\ 
        +2\sqrt{2}(\beta^4-3\beta^2+2) g_s^2 m_t v c_{tG} \cVVuO
        -(2-\beta^2)m_t^2(9 (\cAAuS)^2 +9 (\cAVuS)^2
        + 2((\cAAuO)^2+ 9 (\cAVuO)^2))
    \bigg),
\end{align}
where we note that the quadratic contribution has a pole at $\beta=1$.

\twocolumngrid
\bibliography{apssamp}

\end{document}